\documentclass[twocolumn,astrosymb]{aastex62}

\PassOptionsToPackage{pdftex}{graphicx}
\usepackage{morefloats}
\usepackage{epstopdf}
\usepackage{amsmath}
\usepackage{amssymb}
\usepackage{latexsym}
\usepackage{multirow}
\usepackage{tensor}
\usepackage{enumitem}
\PassOptionsToPackage{pdfpagelabels}{hyperref}
\hypersetup{linkcolor=red,citecolor=blue,filecolor=cyan,urlcolor=teal}
\received{\today}
\revised{N/A}
\accepted{N/A}
\submitjournal{ApJ}
\shorttitle{ELECTRON PARTITION ANALYSIS}
\shortauthors{Wilson III et al.}
\makeatletter
\renewcommand\@makefnmark{\hbox{\@textsuperscript{\normalfont\color{violet}\@thefnmark}}}
\makeatother
\newcommand{\totalnfitsall}{15,210}      
\newcommand{\totalnfitsups}{6546}        
\newcommand{\totalnfitsdns}{8664}        
\newcommand{\totalchbfitsA}{10,983}      
\newcommand{\totalffpfitsA}{12,081}      
\newcommand{\totalshownWHI}{9362}        
\newcommand{\totalncoreall}{14,418}      
\newcommand{\totalnhaloall}{13,660}      
\newcommand{\totalnbeamall}{11,578}      

\begin{document}

\title{Electron energy partition across interplanetary shocks: III.  Analysis}
\correspondingauthor{L.B. Wilson III}
\email{lynn.b.wilsoniii@gmail.com}

\author[0000-0002-4313-1970]{Lynn B. Wilson III}
\affiliation{NASA Goddard Space Flight Center, Heliophysics Science Division, Greenbelt, MD 20771, USA.}

\author[0000-0002-4768-189X]{Li-Jen Chen}
\affiliation{NASA Goddard Space Flight Center, Heliophysics Science Division, Greenbelt, MD 20771, USA.}

\author[0000-0002-6783-7759]{Shan Wang}
\affiliation{Astronomy Department, University of Maryland, College Park, MD 20742, USA.}
\affiliation{NASA Goddard Space Flight Center, Heliophysics Science Division, Greenbelt, MD 20771, USA.}

\author[0000-0003-0682-2753]{Steven J. Schwartz}
\affiliation{Laboratory for Atmospheric and Space Physics, University of Colorado, Boulder, Boulder, CO 80303, USA.}

\author[0000-0002-2425-7818]{Drew L. Turner}
\affiliation{The Johns Hopkins University Applied Physics Laboratory, Laurel, MD 20723, USA.}

\author[0000-0002-7728-0085]{Michael L. Stevens}
\affiliation{Harvard-Smithsonian Center for Astrophysics, Harvard University, Cambridge, MA 02138, USA.}

\author[0000-0002-7077-930X]{Justin C. Kasper}
\affiliation{University of Michigan, Ann Arbor, School of Climate and Space Sciences and Engineering, Ann Arbor, MI 48109, USA.}

\author[0000-0003-2555-5953]{Adnane Osmane}
\affiliation{Department of Physics, University of Helsinki, Helsinki, Finland FI-00014.}

\author[0000-0003-0939-8775]{Damiano Caprioli}
\affiliation{Department of Astronomy and Astrophysics, University of Chicago, Chicago, IL 60637, USA.}

\author[0000-0002-1989-3596]{Stuart D. Bale}
\affiliation{Physics Department, University of California, Berkeley, CA 94720-7300, USA.}
\affiliation{Space Sciences Laboratory, University of California, Berkeley, CA 94720-7450, USA.}
\affiliation{The Blackett Laboratory, Imperial College London, London, SW7 2AZ, UK.}
\affiliation{School of Physics and Astronomy, Queen Mary University of London, London E1 4NS, UK.}

\author[0000-0002-1573-7457]{Marc P. Pulupa}
\affiliation{Space Sciences Laboratory, University of California, Berkeley, CA 94720-7450, USA.}

\author[0000-0002-6536-1531]{Chadi S. Salem}
\affiliation{Space Sciences Laboratory, University of California, Berkeley, CA 94720-7450, USA.}

\author[0000-0002-4288-5084]{Katherine A. Goodrich}
\affiliation{Space Sciences Laboratory, University of California, Berkeley, CA 94720-7450, USA.}

\begin{abstract}
  Analysis of model fit results of \totalnfitsall~electron velocity distribution functions (VDFs), observed within $\pm$2 hours of 52 interplanetary (IP) shocks by the \emph{Wind} spacecraft near 1 AU, is presented as the third and final part on electron VDFs near IP shocks.  The core electrons and protons dominate in the magnitude and change in the partial-to-total thermal pressure ratio, with the core electrons often gaining as much or more than the protons.  Only a moderate positive correlation is observed between the electron temperature and the kinetic energy change across the shock, while weaker, if any, correlations were found with any other macroscopic shock parameter.  No VDF parameter correlated with the shock normal angle.  The electron VDF evolves from a narrowly peaked core with flaring suprathermal tails in the upstream to either a slightly hotter core with steeper tails or much hotter flattop core with even steeper tails downstream of the weaker and strongest shocks, respectively.  Both quasi-static and fluctuating fields are examined as possible mechanisms modifying the VDF but neither is sufficient alone.  For instance, flattop VDFs can be generated by nonlinear ion acoustic wave stochastic acceleration (i.e., inelastic collisions) while other work suggested they result from the combination of quasi-static and fluctuating fields.  This three-part study shows that not only are these systems not thermodynamic in nature, even kinetic models may require modification to include things like inelastic collision operators to properly model electron VDF evolution across shocks or in the solar wind.
\end{abstract}


\phantomsection   
\section{Background and Motivation}  \label{sec:introduction}

\indent  Despite its collisionless, non-equilibrium nature the existence of shock waves is still possible in the solar wind due to solar drivers (e.g., coronal mass ejections and/or corotating interaction regions) called interplanetary (IP) shocks \citep[e.g.,][]{aguilarrodriguez11a, breneman10a, gosling93b, wilsoniii17c}, planetary bow shocks \citep[e.g.,][]{kellogg62a, wilsoniii16a}, cometary bow shocks \citep[e.g.,][]{sagdeev87a}, and from nonlinearly steepened electromagnetic waves radiated by instabilities \citep[e.g.,][]{wilsoniii09a, wilsoniii13a}.  Even though the mean free path of a typical proton near 1 AU can be nearly 1 AU, i.e., orders of magnitude larger than the corresponding thermal gyroradii ($\rho{\scriptstyle_{cp}}$) or inertial length ($\lambda{\scriptstyle_{p}}$), the thickness of shock ramps -- the spatial gradient scale length of the magnetic transition region -- in the solar wind are often a few $\lambda{\scriptstyle_{e}}$ up to $\lambda{\scriptstyle_{p}}$ \citep[e.g.,][]{hobara10a, mazelle10a}.  This is why shock waves in the solar wind, and in most astrophysical contexts, are called collisionless.

\indent  Collisionless shock ramp thickness is thought to depend upon macroscopic shock parameters like the Mach number ($M{\scriptstyle_{f}}$), shock normal angle, $\theta{\scriptstyle_{Bn}}$ (e.g., quasi-perpendicular shocks satisfy $\theta{\scriptstyle_{Bn}}$ $\geq$ 45$^{\circ}$), and upstream averaged plasma beta ($\langle \beta{\scriptstyle_{tot}} \rangle{\scriptstyle_{up}}$).  Therefore, shocks in the solar wind/heliosphere are categorized by these parameters as being low(high) Mach number, $M{\scriptstyle_{f}}$ $\lesssim$ 2.5($M{\scriptstyle_{f}}$ $>$ 2.5); and low(high) beta shocks, $\beta{\scriptstyle_{up}}$ $\leq$ 1.0($\beta{\scriptstyle_{up}}$ $>$ 1.0) \citep[e.g.,][]{coroniti70b, kennel85a, sagdeev66, tidman71a, wilsoniii17c}.

\indent  There are several key unresolved questions about the microphysical processes that regulate the dynamics of collisionless shock waves.  One of the biggest outstanding problems in shock physics is the partition of energy between electrons and ions.  A significant obstacle is that the mechanisms depositing/transferring energy are not predicted to act homogeneously, i.e., they are energy- and pitch-angle-dependent and can be species-dependent \citep[e.g.,][]{artemyev13j, artemyev14e, artemyev15d, artemyev16b, artemyev17a, artemyev17b, artemyev18a, sagdeev66}.  Evidence supporting this prediction have been reported in some case study observations at IP shocks \citep[e.g.,][]{wilsoniii09a, wilsoniii10a, wilsoniii12c, wilsoniii13a} and the terrestrial bow shock \citep[e.g.,][]{chen18a, goodrich18c, goodrich19a, oka17a, oka19a, wilsoniii14a, wilsoniii14b}.  Further, most collisionless shocks are subsonic to electrons, yet electrons still behave as if they have experienced a shock, even showing Mach number dependent effects \citep[e.g.,][]{feldman82a, feldman83a, feldman83b, masters11a, thomsen85a, thomsen87b, thomsen93a, wilsoniii10a}.  The question remains, how does a collisionless shock transform the bulk flow kinetic energy into other forms like electron and/or ion heating.

\indent  To determine the energy transfer mechanism, several previous studies examined the change in average electron temperature, $\Delta \bar{T}{\scriptstyle_{e, tot}}$ (see Appendix \ref{app:Definitions} for definitions), or the ratio of downstream-to-upstream average electron temperature, $\bar{R}{\scriptstyle_{Te,tot}}$, across the shock.  These parameters were compared with the upstream average fast Mach number, $\langle M{\scriptstyle_{f}} \rangle{\scriptstyle_{up}}$, the upstream average kinetic energy, $\langle \lvert KE{\scriptstyle_{shn}} \rvert \rangle{\scriptstyle_{up}}$, the change in the shock normal speed, $\Delta \bar{U}{\scriptstyle_{shn}}$, and the square of the shock normal speed, $\Delta \bar{U}{\scriptstyle_{shn}}^{2}$ and/or the change in kinetic energy across the shock\footnote{Some studies explicitly calculated the change in kinetic energy across the shock, $\Delta \overline{KE}{\scriptstyle_{shn}}$, while others only calculated $\Delta \bar{U}{\scriptstyle_{shn}}^{2}$.}.

\indent  \citet[][]{feldman83b} found weak, positive correlation between $\bar{R}{\scriptstyle_{Te,tot}}$ and $\langle M{\scriptstyle_{f}} \rangle{\scriptstyle_{up}}$ (technically it was with $\langle M{\scriptstyle_{ms}} \rangle{\scriptstyle_{up}}$).  \citet[][]{feldman83c} found weak, positive linear relationships between both $\Delta \bar{T}{\scriptstyle_{e, tot}}$ and $\Delta \bar{T}{\scriptstyle_{p,tot}}$ and $\Delta \bar{U}{\scriptstyle_{shn}}$.  \citet[][]{thomsen87b} examined electron heating at Earth's bow shock finding a correlation between $\Delta \bar{T}{\scriptstyle_{e, tot}}$ and $\Delta \bar{U}{\scriptstyle_{shn}}^{2}$ with a few to $\sim$10\% of the kinetic energy transforming into electron heating.  Later \citet[][]{schwartz88a} found a correlation between $\Delta \bar{T}{\scriptstyle_{e, tot}}$ and $\Delta \bar{U}{\scriptstyle_{shn}}^{2}$.  They also show a strong, positive correlation for $\Delta \bar{T}{\scriptstyle_{e, tot}}$ versus $\Delta \bar{T}{\scriptstyle_{i, tot}}$ with a slope of $\sim$0.2 and y-intercept of $\sim$8.6 eV.  \citet[][]{thomsen93a} examined ion and electron heating at the low Mach number, quasi-parallel Earth's bow shock finding that $\sim$6\% of $\Delta \overline{KE}{\scriptstyle_{shn}}$ was converted to electron heating.  \citet[][]{hull00a} found a positive, linear relationship between $\Delta \bar{T}{\scriptstyle_{e, tot}}$ and $\Delta \overline{KE}{\scriptstyle_{shn}}$ with a slope of $0.057 \ \substack{+0.0041 \\ -0.0039}$, i.e., $\sim$6\% of of $\Delta \overline{KE}{\scriptstyle_{shn}}$ was converted to electron heating.  \citet[][]{fitzenreiter03a} also found a linear relationship between $\Delta \bar{T}{\scriptstyle_{e, tot}}$ and $\Delta \bar{U}{\scriptstyle_{shn}}^{2}$.  Finally, \citet[][]{masters11a} examined electron heating at the Kronian bow shock finding that for most crossings $\sim$3\%--7\% of $\langle KE{\scriptstyle_{shn}} \rangle{\scriptstyle_{up}}$ was converted to electron heating.  Thus, previous work suggests that the change in bulk flow kinetic energy should be correlated with the increase in temperature.  Note that previous work has either inferred \citep[e.g.,][]{ghavamian07a, ghavamian14a} or showed with in situ measurements that stronger shocks heat ions more than electrons \citep[e.g.,][]{schwartz88a, thomsen93a, masters11a} with the fraction of electron heating varying with an inverse Mach number\footnote{However, more complete analysis of astrophysical shocks that include neutral return currents found that the electron-to-ion temperature ratio cannot be constrained by Mach number \citep[e.g.,][]{blasi12a}.}.  The inverse Mach number dependence was also found to be a piecewise distribution where $\sim$50\% of the total heating goes to electrons at low Mach number and changes to $\lesssim$10\% at higher Mach number.

\indent  In this third part (Paper III) of this three-part study, the analysis of the fit results to the multi-component electron VDF analysis will be discussed in the context of the macroscopic shock parameters and some instability analysis.  The results are summarized for the 52 IP shocks observed by the \emph{Wind} spacecraft.  The notation, symbols, and data sets used herein are the same as those in \citet[][]{wilsoniii19a} (hereafter referred to as Paper I) and \citet[][]{wilsoniii19b} (hereafter referred to as Paper II).  Paper I discussed the methodology and described the data product resulting from the application of the fit software and Paper II summarized the statistics of the fit results.  This work will provide the physical analysis and interpretation of the results in the context of macroscopic shock parameters and some instability analysis.

\section{Data Sets and Methodology}  \label{sec:DefinitionsDataSets}

\indent  As in Papers I and II, all data are observed by instruments on the \emph{Wind} spacecraft \citep{harten95a} near 1 AU.  The data used herein include quasi-static magnetic field vectors ($\mathbf{B}{\scriptstyle_{o}}$) from \emph{Wind}/MFI \citep[][]{lepping95}, electron and ion velocity distribution functions (VDFs) from \emph{Wind}/3DP \citep[][]{lin95a}, and proton and alpha-particle velocity moments from the \emph{Wind}/SWE Faraday Cups \citep[][]{kasper06a, ogilvie95}.  The instrument details are described in Paper I.  Parameters described with respect to $\mathbf{B}{\scriptstyle_{o}}$ are in a field-aligned coordinate basis using a subscript $j$ to denote the parallel ($j$ $=$ $\parallel$), the perpendicular ($j$ $=$ $\perp$), and total ($j$ $=$ $tot$) directions.  All electron parameters are shown with a subscript $s$ denoting the component (or sub-population) of the entire distribution where $s$ $=$ $ec$ for the core, $s$ $=$ $eh$ for the halo, $s$ $=$ $eb$ for the beam/strahl, and $s$ $=$ $e$ for the entire distribution.  The combined or mixed parameters (e.g., $\beta{\scriptstyle_{eff, j}}$) use the subscripts $s$ $=$ $eff$ for \emph{effective} and $s$ $=$ $int$ for \emph{integrated} parameters (see Appendix \ref{app:Definitions} for definitions).  The analysis of the VDFs presented herein were found within $\pm$2 hours of 52 IP shocks found in the \emph{Wind} shock database from the Harvard Smithsonian Center for Astrophysics\footnote{\url{https://www.cfa.harvard.edu/shocks/wi\_data/}} between 1995-02-26 and 2000-02-20 (for full list of event dates and times, see PDF file included with additional supplemental material found at \url{https://doi.org/10.5281/zenodo.2875806} \citet[][]{wilsoniii19k}).  The IP shocks examined were limited to fast-forward shocks that had burst mode electron VDFs within the chosen time range about each shock.

\indent  As in Paper II, the VDF results presented herein relative to all \totalnfitsall~VDFs examined.  The VDF fit results are taken from additional supplemental material in the form of two ASCII files (discussed in Paper II and described in Paper I) \citep[][]{wilsoniii19k}.  Of this total \totalncoreall~had stable model fits ($f^{\left( core \right)}$) for the core, \totalnhaloall~stable model fits ($f^{\left( halo \right)}$) for the halo, and \totalnbeamall~stable model fits ($f^{\left( beam \right)}$) for the beam/strahl.  Note that all statistics presented herein are for stable fits with a fit flag for the respective component of two or higher.  The selection justification of the VDFs are provided in Paper II.  Similarly, this work follows Paper II with the following selection criteria:
\begin{itemize}[itemsep=0pt,parsep=0pt,topsep=0pt]
  \item[] \textit{Criteria AT:} All VDFs satisfying: Fit Flag $\{c,h,b\}$ $\geq$ 2 and no violation of post-fit constraints;
  \item[] \textit{Criteria UP:} All VDFs satisfying \textit{Criteria AT} that were observed upstream of the IP shock ramp;
  \item[] \textit{Criteria DN:} All VDFs satisfying \textit{Criteria AT} that were observed downstream of the IP shock ramp;
  \item[] \textit{Criteria LM:} All VDFs satisfying \textit{Criteria AT} that were observed near IP shocks satisfying $\langle M{\scriptstyle_{f}} \rangle{\scriptstyle_{up}}$ $<$ 3;
  \item[] \textit{Criteria HM:} All VDFs satisfying \textit{Criteria AT} that were observed near IP shocks satisfying $\langle M{\scriptstyle_{f}} \rangle{\scriptstyle_{up}}$ $\geq$ 3;
  \item[] \textit{Criteria PE:} All VDFs satisfying \textit{Criteria AT} that were observed near IP shocks satisfying $\theta{\scriptstyle_{Bn}}$ $>$ 45$^{\circ}$; and
  \item[] \textit{Criteria PA:} All VDFs satisfying \textit{Criteria AT} that were observed near IP shocks satisfying $\theta{\scriptstyle_{Bn}}$ $\leq$ 45$^{\circ}$.
\end{itemize}

\indent  Superposed epoch analysis (SEA) of every fit parameter has been performed where the time stamps of each VDF are redefined as offsets from the associated IP shock ramp center time.  Finally, to quantify the partition of energy for each component, all permutations of the difference (and ratio) between the upstream and downstream values are computed.  Again, the value used for each shock is the median and the uncertainty is half the magnitude of the difference between $X{\scriptstyle_{5\%}}$ and $X{\scriptstyle_{95\%}}$.  The results were fit to model functions (e.g., power-law) with the dependent variable/abscissa being the different macroscopic shock parameters.  All permutations are used to avoid the subjectivity introduced by selecting a user-defined time period that serves as the upstream and downstream regions.  The median was used instead of an average because again the data are normally distributed and the median is less influenced by large deviations/tails.  The differences or ratios are then fit to a model function using both Poisson and Gaussian weights to find the best fit results.

\indent  Finally, it should be noted that the total time range (i.e., $\pm$2 hours) about each shock ramp is larger than the typical time range examined (i.e., $\pm$10s of minutes).  The purpose is ensure that the upstream analysis includes data that is not part of the electron foreshock.  The extent of the downstream is defined purely to maintain symmetry about the shock ramp center.  Although the analysis presented herein only discusses differences and analysis of the VDF solutions for the entire time range, the same analysis was performed on limited time ranges about the shock ramp for comparison (not shown).  Alteration of the time range about the shock ramp did not yield significant differences for most parameters examined.

\phantomsection   
\section{Superposed Epoch Analysis}  \label{sec:SuperposedEpochAnalysis}

\indent  In the following subsections multiple figures presenting superposed epoch analysis (SEA) of the various fit parameters are presented.  In all sections, the subscripts $s$ $=$ $ec$ is for the core, $s$ $=$ $eh$ for the halo, $s$ $=$ $eb$ for the beam/strahl, and $s$ $=$ $eff$ for the effective components.

\indent  The time stamps of each VDF are redefined as offsets from the associated IP shock ramp center time.  The data are partitioned into 120 second time windows and then one-variable statistical analysis is performed on all data within.  The time window center is shifted by 22 seconds (i.e., roughly the sample period of the 3DP instrument in survey mode) and then the analysis is repeated.  In this way, the data can remain at their original sample time without affecting the magnitudes through the use of a re-gridding algorithm by averaging or interpolating data to a common set of time stamps.

\indent  For every partitioned time window, $\tilde{X}$ is used as the center line (shown as a red line in Figures \ref{fig:Exponents} and \ref{fig:DensRatios}), $X{\scriptstyle_{5\%}}$ as the lower line (shown as the lower cyan line in Figures \ref{fig:Exponents} and \ref{fig:DensRatios}), and $X{\scriptstyle_{95\%}}$ as the upper line (shown as the upper cyan line in Figures \ref{fig:Exponents} and \ref{fig:DensRatios}).  The median and percentile lines are then smoothed by selection criteria-dependent widths\footnote{As shown in Paper II, there are fewer fit results for \textit{Criteria HM} than \textit{Criteria LM} and so larger smoothing windows were required.}, with \textit{Criteria AT} smoothed over 10pts, \textit{Criteria LM} smoothed over 20pts, \textit{Criteria HM} smoothed over 30pts, \textit{Criteria PA} smoothed over 30pts, and \textit{Criteria PE} smoothed over 20pts.  The data are also normalized by the upstream median values for each IP shock (see Tables 2--6 of additional supplemental material found in \citet[][]{wilsoniii20b} for list of values) to remove relative offsets between any two IP shocks.

\indent  The median and percentiles are used as the data are not normally distributed about a mean for each partitioned time window, as implied by the non-Gaussian histogram distributions shown in Paper II.  The SEA plots provide the trend and typical ranges of the various parameters relative to the shock ramp center to illustrate changes.  The parameters are also separated by the above selection criteria to illustrate macroscopic shock parameter dependence.

\phantomsection   
\subsection{Exponents}  \label{subsec:Exponents}

\begin{figure}
  \centering
    {\includegraphics[trim = 0mm 0mm 0mm 0mm, clip, width=80mm]{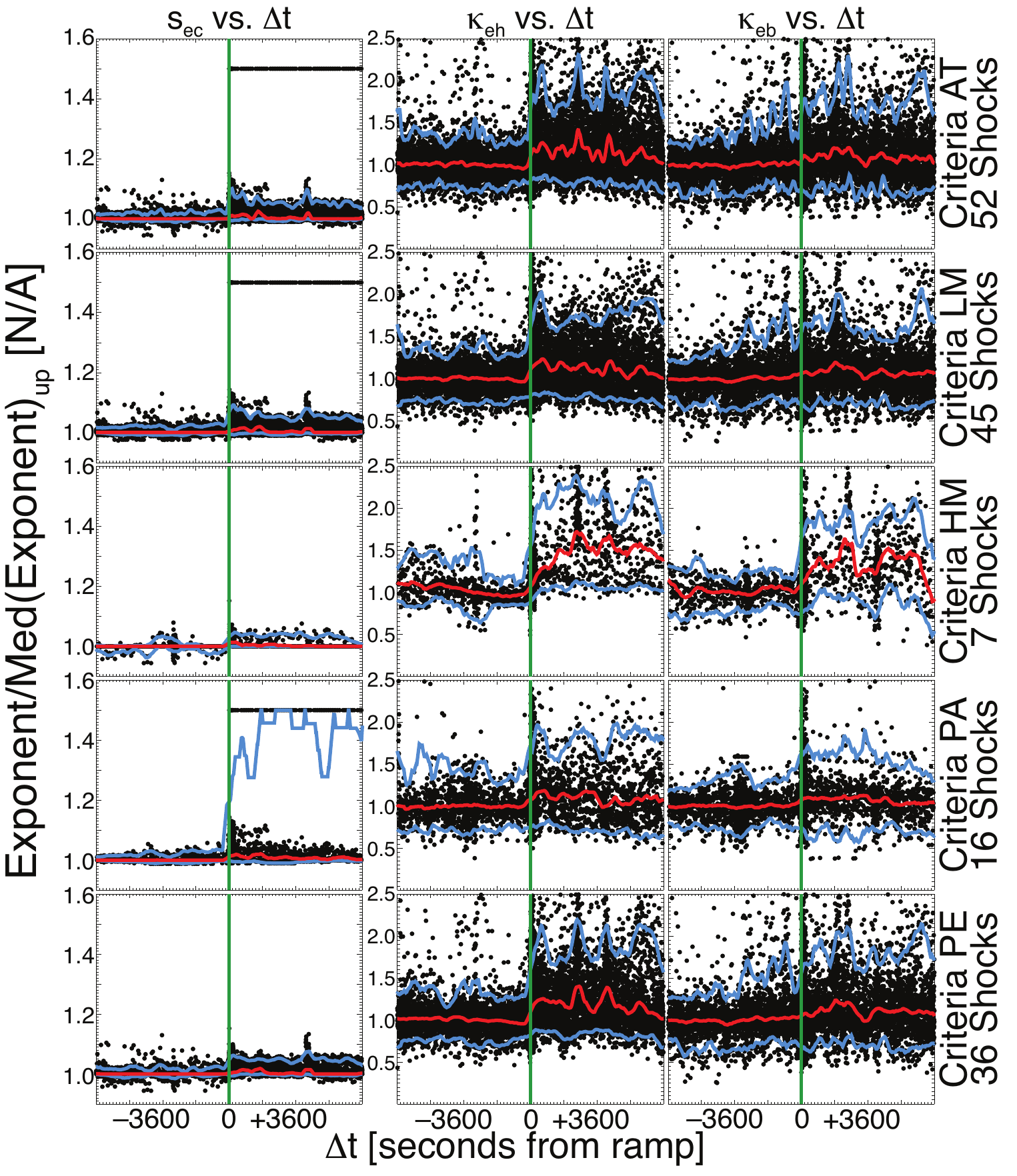}}
    \caption{Superposed epoch analysis plot of the core, $s{\scriptstyle_{ec}}$, halo, $\kappa{\scriptstyle_{eh}}$ (middle column), and beam/strahl, $\kappa{\scriptstyle_{eb}}$ (right column), exponents separated into \textit{Criteria AT} (first row), \textit{Criteria LM} (second row), \textit{Criteria HM} (third row), \textit{Criteria PA} (fourth row), \textit{Criteria PE} (fifth row).  The red line shows the smoothed $\tilde{X}$ values of the partitioned data.  The lower(upper) cyan lines are the smoothed $X{\scriptstyle_{5\%}}$($X{\scriptstyle_{95\%}}$) values of the partitioned data.  The vertical green line indicates the ramp center time.  All data for a given IP shock are normalized by an upstream median value given in Table 5 found in \citet[][]{wilsoniii20b}.  Note that all panels share the same horizontal axis range but the first column has a different vertical range than the latter two.}
    \label{fig:Exponents}
\end{figure}

\indent  In this section, SEAs of $s{\scriptstyle_{ec}}$/$\langle s{\scriptstyle_{ec}} \rangle{\scriptstyle_{up}}$, $\kappa{\scriptstyle_{eh}}$/$\langle \kappa{\scriptstyle_{eh}} \rangle{\scriptstyle_{up}}$, and $\kappa{\scriptstyle_{eb}}$/$\langle \kappa{\scriptstyle_{eb}} \rangle{\scriptstyle_{up}}$ are introduced and discussed in Figure \ref{fig:Exponents} (see Appendix \ref{app:Definitions} for definitions).  First, 50\% of the halo and beam/strahl exponents (i.e., between the lower and upper quartiles) satisfy 3.57 $\lesssim$ $\kappa{\scriptstyle_{eh}}$ $\lesssim$ 5.31 and 3.41 $\lesssim$ $\kappa{\scriptstyle_{eb}}$ $\lesssim$ 5.11, which are consistent with previous solar wind observations near 1 AU \citep[e.g.,][]{horaites18a, lazar17a, maksimovic97a, maksimovic05a, pierrard16a, stverak09a, tao16a, tao16b} (there are no studies with which to compare the $s{\scriptstyle_{ec}}$ values).

\indent  The $s{\scriptstyle_{ec}}$ exponents are relatively constant across most IP shocks, with a few exceptions for \textit{Criteria LM} and \textit{Criteria PA} shocks showing values up near $\sim$1.5.  The \textit{Criteria HM} are especially sparsely populated in the downstream as many of these VDFs were fit to the asymmetric self-similar model distribution (i.e., $p{\scriptstyle_{ec}}$ and $q{\scriptstyle_{ec}}$ exponents, not shown).  However, the range of data between $X{\scriptstyle_{5\%}}$ and $X{\scriptstyle_{95\%}}$ is larger in the downstream, consistent with the interpretation of a wave-induced inelastic interaction discussed in Paper I.

\indent  The $\kappa{\scriptstyle_{eh}}$ panels (middle column) are more interesting and varied. For instance, the running median clearly increases across the shock.  The change is the largest for \textit{Criteria HM} and \textit{Criteria PE} shocks.  Theory and simulation suggest that stronger and more oblique shocks are better at energizing suprathermal electrons \citep[e.g.,][]{caprioli14a, park15a, treumann09a, trotta19a}.  However, if the normalization used in Figure \ref{fig:Exponents} is removed (not shown), the upstream only values of $\kappa{\scriptstyle_{eh}}$ between $X{\scriptstyle_{min}}$ and $X{\scriptstyle_{max}}$ cover a similar range for \textit{Criteria LM} and \textit{Criteria HM} shocks, with similar running medians as well.  The downstream only $\kappa{\scriptstyle_{eh}}$ values are larger at \textit{Criteria HM} than \textit{Criteria LM} shocks, which explains the larger one-variable statistics values of $\kappa{\scriptstyle_{eh}}$ for \textit{Criteria HM} than \textit{Criteria LM} shocks shown in Paper I.

\indent  Note that the one-variable statistics of $\kappa{\scriptstyle_{eh}}$ reported in Paper I showed a slightly larger value for \textit{Criteria PE} than \textit{Criteria PA} shocks in agreement with the larger change across \textit{Criteria PE} than \textit{Criteria PA} shocks observed in Figure \ref{fig:Exponents}, consistent with theory and simulations \citep[e.g.,][]{wu84b, park13a, trotta19a}.  While the change is not in the direction expected, the magnitude of the change is certainly larger for \textit{Criteria HM} than \textit{Criteria LM} shocks suggesting stronger shocks have more of an impact on the halo than weaker.  While the upstream kappa values have recently been predicted to increase with increasing Mach number for quasi-perpendicular shocks \citep[][]{trotta19a}, the normalization by the upstream median values for each shock removed the relative offsets to avoid this potentially misleading difference.

\indent  The increase in $\kappa{\scriptstyle_{eh}}$ across the shocks could result from several effects.  For instance, it could result from core electrons being energized to suprathermal energies, initially starting with a larger equivalent kappa value, thus increasing the overall suprathermal exponent value.  The increase in $\kappa{\scriptstyle_{eh}}$ could also result from acceleration through a quasi-static electric field \citep[e.g.,][]{mitchell14a, schwartz88a, schwartz11a, schwartz14a, scudder86c} for a one-dimensional VDF.  The change in $\kappa{\scriptstyle_{eh}}$ for a two-dimensional VDF in velocity space is complicated if the quasi-static electric field is not uniformly aligned with both component directions in velocity space, as a kappa VDF is not a simple power-law \citep[e.g., ][]{livadiotis15a}.  Even so, the effect of a quasi-static electric field could result in the fit software finding a larger kappa value as the higher energy particles change velocity by a smaller fraction than the lower.  However, the nonlinear relationship between the values of $\kappa{\scriptstyle_{s}}$, $n{\scriptstyle_{s}}$, and $V{\scriptstyle_{Ts, j}}$ in affecting the shape and peak phase space density of a bi-kappa VDF (e.g., see Figure 2 of Paper I) makes it difficult to interpret the change in $\kappa{\scriptstyle_{eh}}$.

\indent  The $\kappa{\scriptstyle_{eb}}$ panels (right-hand column) are less dynamic than the $\kappa{\scriptstyle_{eh}}$ panels but there is important similarity.  The change in the running median is largest for \textit{Criteria HM} shocks, with all other selection criteria showing little-to-no change or a weak gradual change across the entire time range.  The biggest difference between $\kappa{\scriptstyle_{eh}}$ and $\kappa{\scriptstyle_{eb}}$ is that the latter does not show a clear increase at the shock ramp for \textit{Criteria PE} shocks.  Similar to the discussion above for $\kappa{\scriptstyle_{eh}}$, if the normalization is removed from Figure \ref{fig:Exponents} for $\kappa{\scriptstyle_{eb}}$, the difference between \textit{Criteria LM} and \textit{Criteria HM} shocks is mostly in the variation in the running median.  For \textit{Criteria LM} shocks, the running median of $\kappa{\scriptstyle_{eb}}$ is very smooth and does not change much at the shock ramp (i.e., changes by less than $\sim$0.5) while the running median for \textit{Criteria HM} shocks fluctuates with a normalized amplitude at or above unity.  Again, the change in the running median does not show a strong change at the shock ramp in the unnormalized SEA (not shown).

\phantomsection   
\subsection{Density Ratios}  \label{subsec:DensRatios}

\begin{figure}
  \centering
    {\includegraphics[trim = 0mm 0mm 0mm 0mm, clip, width=80mm]{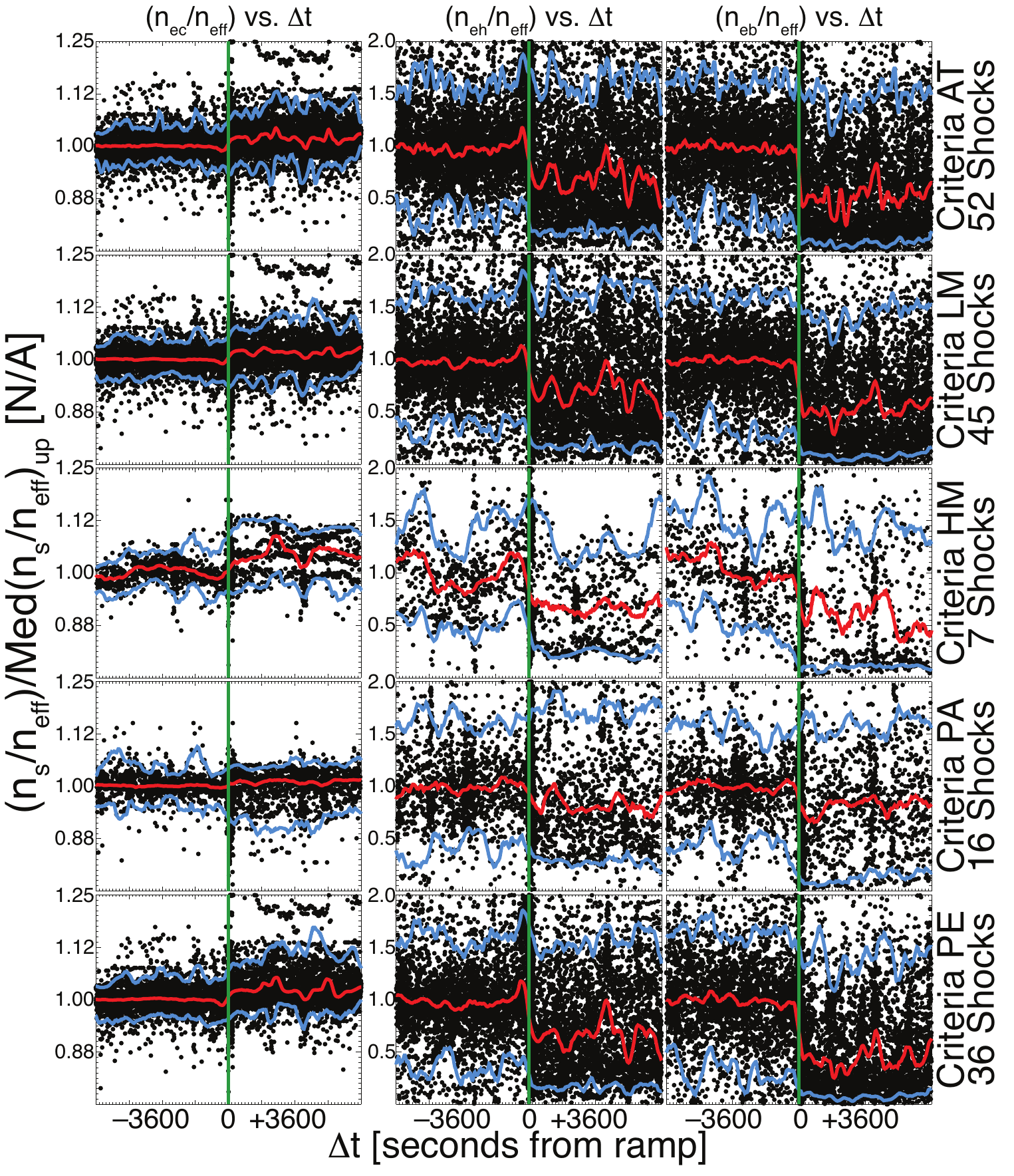}}
    \caption{Superposed epoch analysis plot of the electron component density ratios $n{\scriptstyle_{ec}} / n{\scriptstyle_{eff}}$ (left column), $n{\scriptstyle_{eh}} / n{\scriptstyle_{eff}}$ (middle column), and $n{\scriptstyle_{eb}} / n{\scriptstyle_{eff}}$ (right column).  The normalization values for each IP shock are given in Table 3 found in \citet[][]{wilsoniii20b}.  The format is the same as Figure \ref{fig:Exponents}.}
    \label{fig:DensRatios}
\end{figure}

\indent  In this section, superposed epoch analyses (SEAs) of $n{\scriptstyle_{s}} / n{\scriptstyle_{eff}}$ are introduced and discussed, for the core ($s$ $=$ $ec$), halo ($s$ $=$ $eh$), beam/strahl ($s$ $=$ $eb$), and effective ($s$ $=$ $eff$).  The halo-to-effective density ratios are presented in the first column of Figure \ref{fig:DensRatios}, the halo-to-effective ratios in the second column, and the beam-to-effective ratios in the third column.

\indent  Figure \ref{fig:DensRatios} shows the ratio of each electron component number density to the effective number density, $n{\scriptstyle_{eff}}$ (see Appendix \ref{app:Definitions} for definitions), to illustrate the relative change in the fractional composition of the electron distribution.  That is, the SEA plots show whether the core, halo, and/or beam/strahl number density fraction increase or decrease across the shock for each selection criteria.  It's quite obvious that the median core fraction always increases across the shock, regardless of selection criteria.  The difference between selection criteria is that magnitude of the change, e.g., \textit{Criteria PA} shows very little positive change in the running median and a skewness toward smaller values in the running 5$^{th}$ percentile.

\indent  Unlike the core, both the halo and beam/strahl fractional densities decrease across the shock for all selection criteria.  Again, the weakest change is found in the \textit{Criteria PA} shocks, which is not surprising as quasi-parallel shocks show smaller magnetic field compression ratios.  Thus, the change in the core density is smaller which impacts the halo and beam/strahl fits.  The largest change across the shock ramp occurred for \textit{Criteria LM} and \textit{Criteria PE}.  It should also be noted that the gradient in the running median is sharpest for the beam/strahl component.  The halo shows a short enhancement upstream of the shock ramp and then a decrease that continues shortly into the downstream past the ramp.  The width of the upstream enhancement is larger for the \textit{Criteria HM} than \textit{Criteria LM} shocks, suggesting it results from some form of shock acceleration forming an electron foreshock.  This is further evidenced by the stark difference in the running median profile between the \textit{Criteria PA} and \textit{Criteria PE} shocks, where the latter are expected to be better at energizing electrons \citep[e.g.,][]{wu84b, park13a, trotta19a}.

\indent  The beam/strahl gradients occur almost entirely in the thin region focused on the ramp.  On the time scale shown in these figures, one minor tick mark is $\sim$180 seconds so the halo gradient from upstream to downstream lasts upwards of $\sim$15 minutes.  The beam/strahl gradient is much shorter at $\sim$6 minutes (except for \textit{Criteria PA} and \textit{Criteria HM} shocks which have a much gentler gradient).  Recall the time windows for calculating the running median are 120 seconds and the windows are shifted by 22 seconds each time, thus the sharpest gradient one might expect would be a little over two minutes.

\indent  These plots also show that the fraction of suprathermal particles decreases across most IP shocks, which may explain why the suprathermal exponents tend to increase across the shocks.  Although the suprathermal temperatures tend toward larger values in the downstream regions (see additional SEA plots of $T{\scriptstyle_{s, j}}$ found in \citet[][]{wilsoniii20b}), the change is very weak except for \textit{Criteria HM} shocks.  However, the core temperature changes across the shocks are much more dramatic than both the halo and beam/strahl.  Further, the beam/strahl shows the weakest changes in $T{\scriptstyle_{s, j}}$ for all selection criteria, i.e., the beam/strahl component's mean kinetic energy is not strongly affected by the shock.

\phantomsection   
\subsection{Betas}  \label{subsec:Betas}

\begin{figure}
  \centering
    {\includegraphics[trim = 0mm 0mm 0mm 0mm, clip, width=80mm]{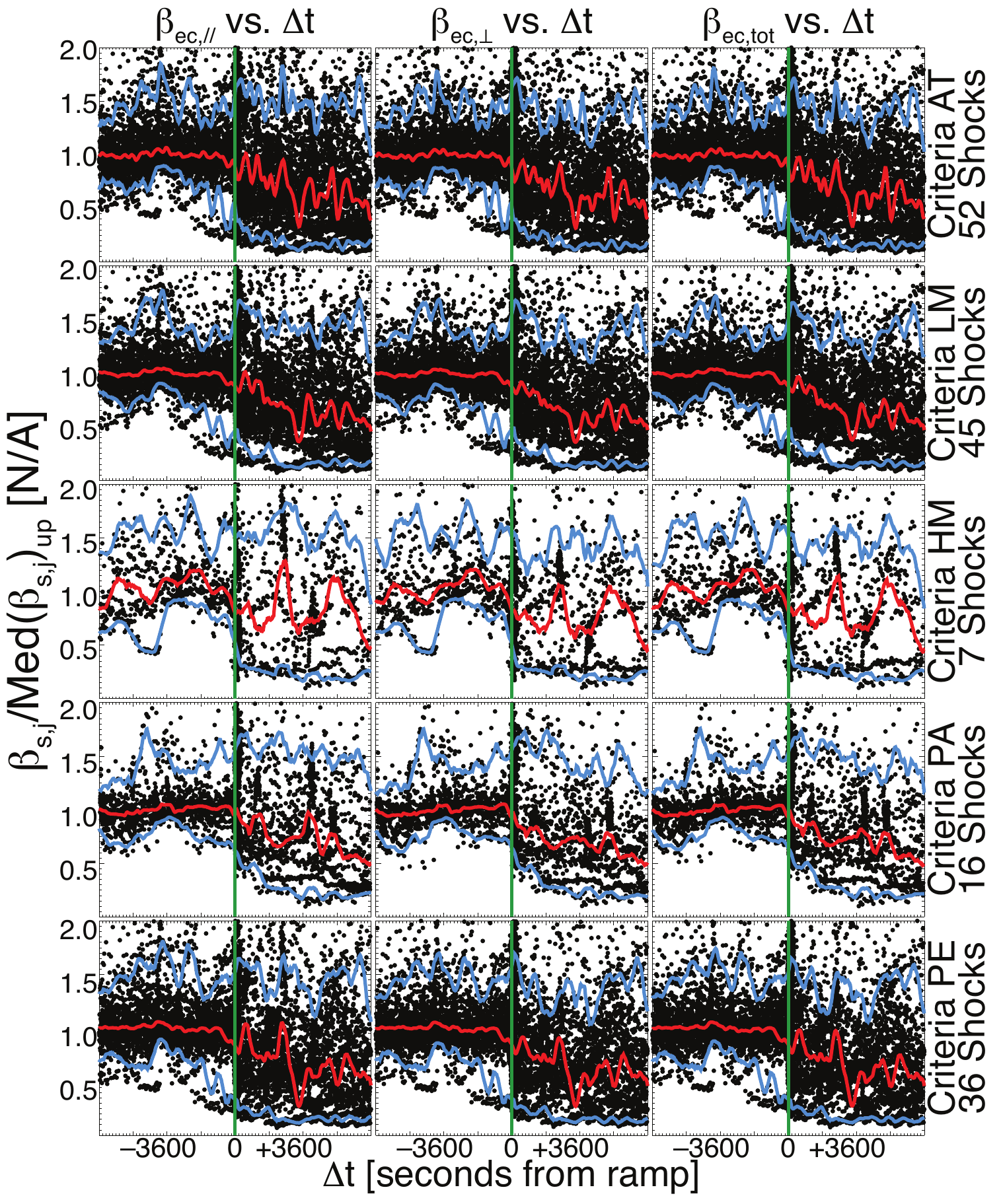}}
    \caption{SEA plots of the core plasma betas $\beta{\scriptstyle_{ec, \parallel}}$ (left column), $\beta{\scriptstyle_{ec, \perp}}$ (middle column), and $\beta{\scriptstyle_{ec, tot}}$ (right column).  The normalization values for each IP shock are given in Table 4 found in \citet[][]{wilsoniii20b}.  The format is similar Figure \ref{fig:Exponents} but all three columns share the same vertical axis scale.}
    \label{fig:CoreBeta}
\end{figure}

\begin{figure}
  \centering
    {\includegraphics[trim = 0mm 0mm 0mm 0mm, clip, width=80mm]{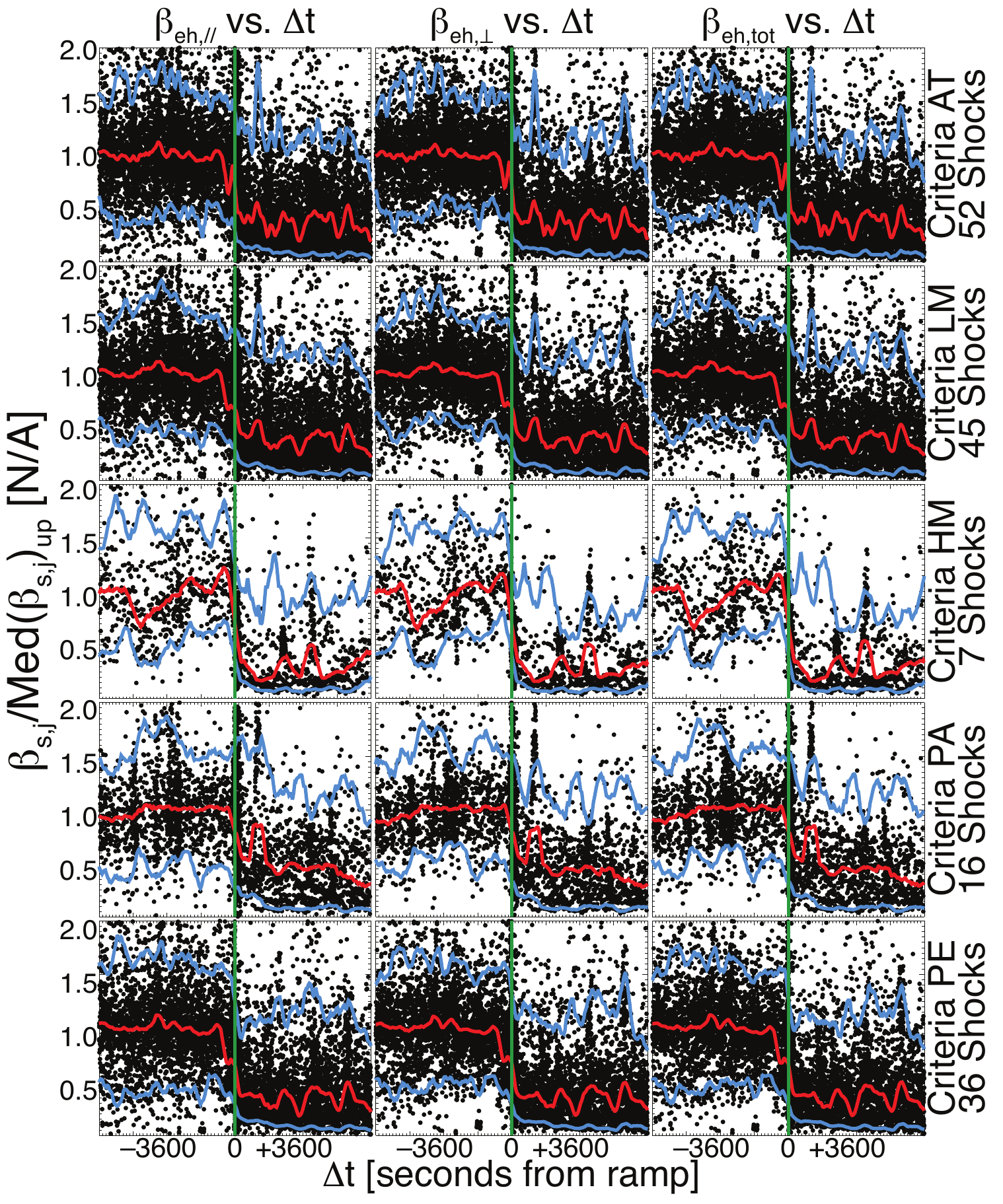}}
    \caption{SEA plots of the halo plasma betas $\beta{\scriptstyle_{eh, \parallel}}$ (left column), $\beta{\scriptstyle_{eh, \perp}}$ (middle column), and $\beta{\scriptstyle_{eh, tot}}$ (right column).  The normalization values for each IP shock are given in Table 4 found in \citet[][]{wilsoniii20b}.  The format is the same as Figure \ref{fig:CoreBeta}.}
    \label{fig:HaloBeta}
\end{figure}

\begin{figure}
  \centering
    {\includegraphics[trim = 0mm 0mm 0mm 0mm, clip, width=80mm]{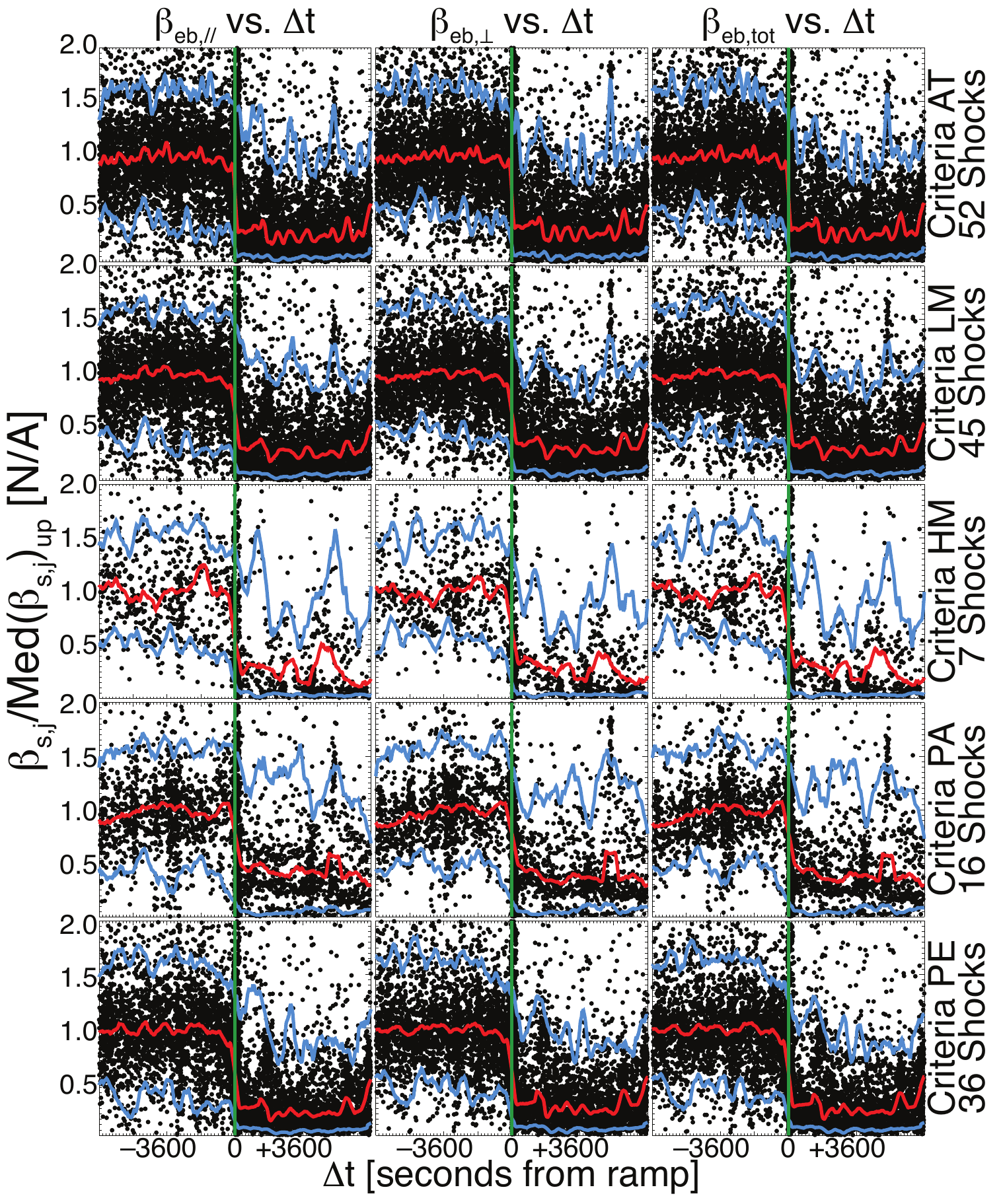}}
    \caption{SEA plots of the beam/strahl plasma betas $\beta{\scriptstyle_{eb, \parallel}}$ (left column), $\beta{\scriptstyle_{eb, \perp}}$ (middle column), and $\beta{\scriptstyle_{eb, tot}}$ (right column).  The normalization values for each IP shock are given in Table 4 found in \citet[][]{wilsoniii20b}.  The format is the same as Figure \ref{fig:CoreBeta}.}
    \label{fig:BeamBeta}
\end{figure}

\indent  In this section, superposed epoch analyses (SEAs) of the electron plasma betas, $\beta{\scriptstyle_{s, j}}$,  for the core ($s$ $=$ $ec$), halo ($s$ $=$ $eh$), and beam/strahl ($s$ $=$ $eb$) components.  The core betas are presented in Figure \ref{fig:CoreBeta}, the halo temperatures in Figure \ref{fig:HaloBeta}, and the beam/strahl temperatures in Figure \ref{fig:BeamBeta}.

\indent  The next thing to examine is the plasma betas, $\beta{\scriptstyle_{s, j}}$,  for the core ($s$ $=$ $ec$), halo ($s$ $=$ $eh$), and beam/strahl ($s$ $=$ $eb$) components.  The SEA plots in Figures \ref{fig:CoreBeta}, \ref{fig:HaloBeta}, and \ref{fig:BeamBeta} of $\beta{\scriptstyle_{s, j}}$ show the largest change across the shock compared to all other electron fit parameters examined.  Of the three components, the core shows the smallest change, but all three electron components decrease across the shock.  The weak decrease in $\beta{\scriptstyle_{ec, j}}$, compared to $\beta{\scriptstyle_{eh, j}}$ and $\beta{\scriptstyle_{eb, j}}$, is dominated by the increase in the magnitude of $\mathbf{B}{\scriptstyle_{o}}$ across the shock since both $T{\scriptstyle_{ec, j}}$ and $n{\scriptstyle_{ec}}$ increase.  Although both $T{\scriptstyle_{eh, j}}$ and $T{\scriptstyle_{eb, j}}$ tend to increase across the shock, the increase is weak.  This weak increase coupled with the stronger decrease in both the fractional $n{\scriptstyle_{eh}}$ and $n{\scriptstyle_{eb}}$ and increase in the magnitude of $\mathbf{B}{\scriptstyle_{o}}$ makes for the corresponding decrease in suprathermal betas.  Note, however, that the change in $n{\scriptstyle_{eff}}$ and $\mathbf{B}{\scriptstyle_{o}}$ are largely governed by the Rankine-Hugoniot conservation relations while the change in any given $T{\scriptstyle_{s, j}}$ is not.  The change in $T{\scriptstyle_{ec, j}}$ is due to the partition of free energy available due to $\Delta \overline{KE}{\scriptstyle_{shn}}$ $\neq$ 0, which will be discussed in Section \ref{sec:EnergyPartition}.

\indent  Similar to the fractional $n{\scriptstyle_{eh}}$, there is evidence of a foreshock in $\beta{\scriptstyle_{eh, j}}$ shown as a two-step decrease, i.e., the beta begins to drop ahead of the shock ramp, reaches a plateau or slight jump near the ramp, then continues to drop to the downstream values for all selection criteria except \textit{Criteria HM} and \textit{Criteria PA}.  The interesting thing is that the upstream running median is roughly near unity for all $\beta{\scriptstyle_{s, j}}$, by design of course, but the downstream running median drops to as low as $\sim$0.20 for $\beta{\scriptstyle_{eh, j}}$ and $\beta{\scriptstyle_{eb, j}}$.  That is, the running median decreases by upwards of $\sim$80\% across the shock for the suprathermal electrons.  Therefore, the ratio of the thermal-to-magnetic field energy density of the suprathermal electrons is upwards of $\sim$80\% smaller in the downstream than upstream.  Another interesting feature is that $\beta{\scriptstyle_{eh, j}}$ begins to decrease upstream of the ramp center near the times when the fractional $n{\scriptstyle_{eh}}$ increases which appears to be due to a local decrease in $T{\scriptstyle_{eh, j}}$.

\indent  There is some bias in the steepness of the gradient due to the cluster of data near the ramp center of the IP shocks.  This is due to the burst mode trigger of the 3DP instrument, which was designed to capture shock ramps in high time resolution.  Thus, the large spread of values near the ramp center are dominating many of the running median calculations through that time period.  In effect, this is actually reducing the variance in the running median seen in the more sparsely sampled far upstream and downstream regions.  Therefore, it is worth noting to avoid over interpreting these gradients.

\begin{figure*}[!htb]
  \centering
    {\includegraphics[trim = 0mm 0mm 0mm 0mm, clip, width=150mm]{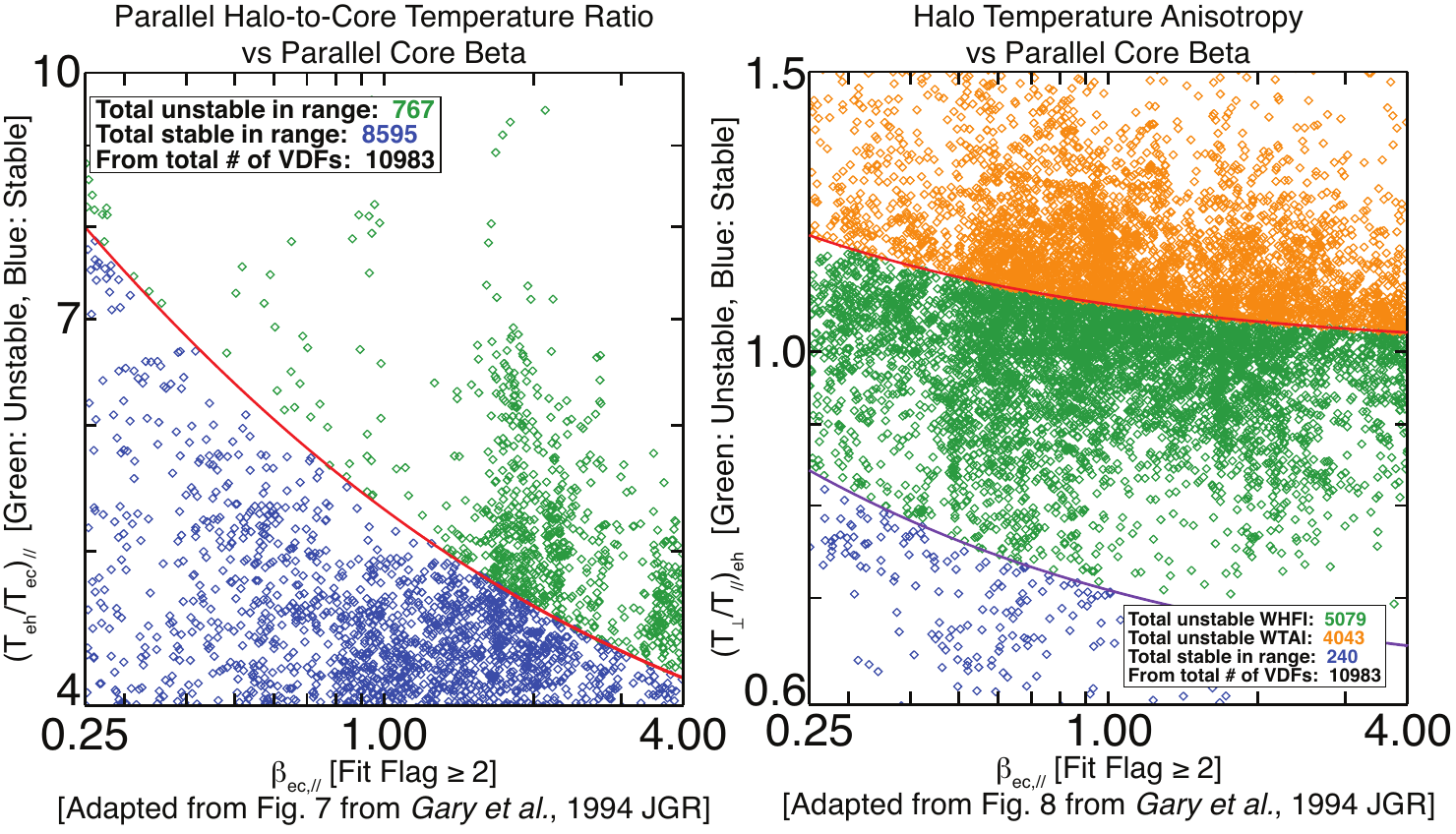}}
    \caption{Adaptations of Figures 7 and 8 from \citet[][]{gary94a} showing the observed data from the current study against the calculated instability threshold for the whistler heat flux (WHFI) and temperature anisotropy (WTAI) instabilities as a function of $\beta{\scriptstyle_{ec, \parallel}}$ for different $\tensor*{ \mathcal{T} }{^{eh}_{ec}}{\scriptstyle_{\parallel}}$ (left-hand panel) and $\mathcal{A}{\scriptstyle_{eh}}$ (right-hand panel) values.  Diamonds shown in green and orange are unstable while blue are stable.  The red and purple lines are the numerical thresholds from the original figures corresponding to $\gamma{\scriptstyle_{max}}$ $>$ 10$^{-1}$ $\Omega{\scriptstyle_{cp}}$.}
    \label{fig:WHFITests}
\end{figure*}

\phantomsection   
\section{Instability Analysis}  \label{sec:InstabilityAnalysis}

\indent  In this section, numerical instability thresholds will be presented.  The purpose of examining various instabilities is driven, in part, by the lack of dependence on many macroscopic shock parameters of the change in many electron fit parameters (this will be discussed in Section \ref{sec:EnergyPartition}) and that waves radiated by instabilities often drive VDFs away from Maxwellians \citep[e.g.,][]{verscharen18a, verscharen19b}, e.g., they generate power-laws and/or other non-Maxwellian features.  There is also the motivation discussed in Paper I regarding the distribution of energy from a wave to the particles.  The issue of inelasticity versus a more standard heating is discussed.  Note that inelastic collisions have been tangentially discussed under different circumstances in previous theoretical work \citep[e.g.,][]{scudder79a}.  That is, inelastic scattering will increase the core exponents, $s{\scriptstyle_{ec}}$ or $p{\scriptstyle_{ec}}$ and $q{\scriptstyle_{ec}}$, but may or may not cause a significant change in $T{\scriptstyle_{ec, j}}$ while the standard idea of heating involves only the change in $T{\scriptstyle_{ec, j}}$ without affecting the exponents.

\indent  Whether the core exponent changes or the temperature is critical for determining the change in the VDF profile/shape.  For instance, whistler modes interacting with an initial Maxwellian VDF can generate strong deviations from Maxwellianity \citep[e.g.,][]{chang13a, gary11a, hughes14a, saito07}, typically resonating with the suprathermal halo and/or beam/strahl \citep[e.g.,][]{coroniti82a, lengyelfrey94, lengyelfrey96, oka17a, oka19a, wilsoniii12c, wilsoniii13a}.  Whistler waves are also of interest because there is some observational evidence that they can generate energetic electron tails \citep[e.g.,][]{oka19a, wilsoniii12c, wilsoniii16h}.  If the whistler is radiated by the whistler heat flux instability (WHFI) \citep[][]{gary94a, gary99a}, during the radiation of the wave fields the skewness of the VDF will reduce and the heat flux carrying electrons will scatter.  The net effect will increase $\mathcal{A}{\scriptstyle_{eb}}$.  The wave fields can also pitch-angle scatter the halo electrons, resulting in larger $\mathcal{A}{\scriptstyle_{eh}}$.  If the whistler is radiated by the temperature anisotropy instability (WTAI), the act of radiating the wave fields will reduce $\mathcal{A}{\scriptstyle_{eh}}$ and/or $\mathcal{A}{\scriptstyle_{eb}}$.  The waves can propagate to another location and pitch-angle scatter the suprathermal particles which increases $\mathcal{A}{\scriptstyle_{eh}}$ and/or $\mathcal{A}{\scriptstyle_{eb}}$ much like the WHFI-driven waves.

\indent  Ion acoustic waves (IAWs) have been observed in the solar wind and near collisionless shocks for over 40 years \citep[e.g.,][]{fredricks68b, fredricks70a, gurnett77a, gurnett79a, kurth79}.  They have been shown to be ubiquitous at collisionless shocks and of large amplitude \citep[e.g.,][]{wilsoniii07a}.  They are thought to be driven by the free energy in currents \citep[e.g.,][]{biskamp72b, lemons78a}, temperature gradients \citep[e.g.,][]{allan74a}, electron heat flux \citep[e.g.,][]{dum80a}, or ion/ion streaming instabilities \citep[e.g.,][]{akimoto85a, akimoto85b, auer71a, goodrich19a} or they can result from a nonlinear wave-wave process \citep[e.g.,][]{cairns92a, dyrud06, kellogg13a, saito17a}.  In the nonlinear stages of their evolution, they have been shown in simulations to cause strong deviations from Maxwellianity toward self-similar core electron VDFs, with $s{\scriptstyle_{ec}}$ or $p{\scriptstyle_{ec}}$ reaching values as large as 5 \citep[e.g.,][]{dum74a, dum75a, dyrud06, sagdeev66, vedenov63}.

\indent  Note that the analysis presented in the following includes a larger than typical time range (i.e., $\pm$2 hours) about the shock ramp.  Therefore, some of the results may be more indicative of the solar wind than the shock itself.  Further, the following analysis uses only the fit parameters from all the electron components and both the protons and alpha-particles \citep[][]{kasper06a} but does not directly include effects of non-Maxwellian features in any of the calculations.  That is, the calculations use fit parameters, like density and temperature, directly in analytical formulas and numerical estimates for a given instability threshold without modification of the formula or numerical estimates.  If the threshold involves only the total ion and electron populations, then the effective electron and total ion parameters are used.

\phantomsection   
\subsection{Whistler Instabilities}  \label{subsec:HeatFluxInstability}

\indent  In this section, the instability thresholds for the WHFI and WTAI are examined based on the observed properties of the electron VDFs.  Specifically, this section will focus on the numerical results of \citet[][]{gary94a} and \citet[][]{gary99a}.  To this end, the electron heat flux is calculated which requires stable fit solutions for all three electron components, i.e., \totalchbfitsA~of the total \totalnfitsall~VDFs examined satisfy this requirement (see Paper II for further details on the integration of the total electron model functions).

\indent  Figure \ref{fig:WHFITests} is an adaptation of Figures 7 and 8 from \citet[][]{gary94a}, where the red(purple) line in the left(right) panel corresponds to the lower line in Figure 7(8) of the original work, i.e., the threshold for the WHFI.  The threshold values shown in Figure \ref{fig:WHFITests} are defined as instabilities having a maximum positive growth rate, $\gamma{\scriptstyle_{max}}$, satisfying $\gamma{\scriptstyle_{max}}$ $>$ 10$^{-1}$ $\Omega{\scriptstyle_{cp}}$.  Over plotted are the electron VDF fit results from the present work.  In the range of $\beta{\scriptstyle_{ec, \parallel}}$ shown, there are \totalshownWHI~valid VDF fit results.  For reference, 90\% of the data for \textit{Criteria AT} satisfy the following:
\begin{itemize}[itemsep=0pt,parsep=0pt,topsep=0pt]
  \item  0.28 rad/s $\lesssim$ $\Omega{\scriptstyle_{cp}}$ $\lesssim$ 2.03 rad/s;
  \item  0.49 s $\lesssim$ $\Omega{\scriptstyle_{cp}}^{-1}$ $\lesssim$ 3.52 s;
  \item  0.76 km $\lesssim$ $\rho{\scriptstyle_{ceff}}$ $\lesssim$ 4.04 km;
  \item  23.6 km $\lesssim$ $\rho{\scriptstyle_{ceff}}$ $\lesssim$ 167 km;
  \item  0.93 km $\lesssim$ $\lambda{\scriptstyle_{ceff}}$ $\lesssim$ 3.65 km; and
  \item  36.7 km $\lesssim$ $\rho{\scriptstyle_{ceff}}$ $\lesssim$ 154 km.
\end{itemize}
\noindent  Thus, the growth times corresponding to the threshold lines in Figure \ref{fig:WHFITests} satisfy 4.92 s $\lesssim$ $\gamma{\scriptstyle_{max}}^{-1}$ $\lesssim$ 35.2 s or $\sim$0.03--0.24\% of the total time examined around each IP shock.

\indent  Figure \ref{fig:WHFITests} shows that $>$80\% of all \totalchbfitsA~VDFs are at or above the threshold for either the WHFI or WTAI.  Limiting to the \totalshownWHI~VDFs shown in Figure \ref{fig:WHFITests}, $\sim$54\% are unstable to the WHFI and $\sim$43\% are unstable to the WTAI.  That is, only $\sim$3\% of the VDFs are stable for these criteria.  Further, \citet[][]{gary99a} noted that in the presence of a finite electron heat flux and $\mathcal{A}{\scriptstyle_{eh}}$ $>$ 1.01, the heat flux carrying electrons are always unstable to the WHFI\footnote{There is the additional criteria that $\beta{\scriptstyle_{ec, \parallel}}$ $>$ 0.25 included in this discussion, but this is already imposed on the data presented in Figure \ref{fig:WHFITests}.}.  Of the \totalchbfitsA~VDFs with a calculated heat flux, $\sim$62\% satisfied $\mathcal{A}{\scriptstyle_{eh}}$ $>$ 1.01, i.e., $\sim$62\% are at least linearly unstable to the WHFI.  These rates are significantly larger than recent work using data from the ARTEMIS mission \citep[e.g.,][]{tong19b}, which found occurrence rates of whistler waves to be $\lesssim$2\%.  However, the ARTEMIS work is limited the observations to the pristine solar wind and used $\mathcal{A}{\scriptstyle_{eff}}$ instead of $\mathcal{A}{\scriptstyle_{eh}}$ for anisotropy threshold calculations.  When they limit their occurrence rate estimates to intervals satisfying $\mathcal{A}{\scriptstyle_{eff}}$ $>$ 1, the rates jump to $\sim$15\%.

\indent  Note that \citet[][]{gary94a} only used two bi-Maxwellian electron components while this work uses three non-Maxwellian electron components, which makes the comparison subject to scrutiny.  For instance, the use of bi-Maxwellian instead of bi-kappa is known to cause differences in the WHFI \citep[e.g.,][]{lee19a, shaaban18a} and others \citep[e.g.,][]{lazar12a, lazar13a, lazar14a, lazar17a, lazar18b, lazar19a, shaaban19a}.  In addition, the definition of $\tensor*{ \mathcal{T} }{^{eh}_{ec}}{\scriptstyle_{\parallel}}$ used to create the left-hand panel in Figure \ref{fig:WHFITests} relied upon the halo and core components only, not a single suprathermal population like that used by \citet[][]{gary94a}.

\indent  More than half of the VDFs are unstable to the WHFI, which is consistent with the observation that both $\mathcal{A}{\scriptstyle_{eb}}$ and $\kappa{\scriptstyle_{eb}}$ are generally smaller than $\mathcal{A}{\scriptstyle_{eh}}$ and $\kappa{\scriptstyle_{eh}}$ (see Paper II for values and statistics), respectively.  That is, the radiation of the wave reduces $\mathcal{A}{\scriptstyle_{eb}}$ and $\kappa{\scriptstyle_{eb}}$ and subsequent scattering can increase $\mathcal{A}{\scriptstyle_{eh}}$.  Though it should be noted that $\kappa{\scriptstyle_{eh}}$ $\gtrsim$ $\kappa{\scriptstyle_{eb}}$ has been found to be true in previous solar wind studies.  That is, statistically the halo exponent tends to be larger than the beam/strahl at 1 AU in some studies \citep[e.g.,][]{stverak09a}, though others show the converse and a solar cycle dependence \citep[e.g.,][]{tao16a}.  Therefore, the larger $\kappa{\scriptstyle_{eh}}$ may be a remnant of the typical conditions in the solar wind and not directly related to local instabilities.  Or the differences between the two exponents may be observed with this relationship precisely because the two populations are intrinsically coupled through whistler-like instabilities.  It may also be that quasi-static fields are affecting the core more than the halo by energizing them to suprathermal energies, thus effectively increasing the suprathermal electron exponents \citep[e.g.,][]{mitchell14a, schwartz88a, schwartz11a, schwartz14a, scudder86c}.  Kinetic simulations are required to resolve this discrepancy and are beyond the scope of this study.

\phantomsection   
\subsection{Other Instabilities}  \label{subsec:OtherInstabilities}

\indent  In this section, the instability thresholds for short wavelength\footnote{Short here implies $\lambda$/$\lambda{\scriptstyle_{De}}$ $\sim$ 2$\pi$ to $>$several 10s, i.e., not the large wavelength limit sometimes called the slow ion acoustic mode.}, electrostatic ion acoustic waves (IAWs) will be calculated and discussed.  The purpose is to calculate the probability of occurrence and determine whether such modes could be playing a role in the evolution of the electron VDFs across the examined IP shocks.

\indent  A long standing problem in solar wind physics is the occurrence of short wavelength IAWs \citep[e.g.,][]{fuselier84a, gurnett79a, gurnett79b, wilsoniii07a} despite the commonly observed electron-to-ion temperature ratios satisfying $\tensor*{ \mathcal{T} }{^{eff}_{i}}{\scriptstyle_{tot}}$ $<$ 3.  The temperature ratio threshold derives from the assumption of single, isotropic Maxwellian VDFs for both the electrons and ions which shows that current-driven IAWs are heavily Landau damped if $\tensor*{ \mathcal{T} }{^{eff}_{i}}{\scriptstyle_{tot}}$ $<$ 3 \citep[e.g.,][]{fried61a, fried66a, gould64a, gurnett79b}.  However, temperature gradients \citep[e.g.,][]{allan74a, dum78a, dum78b, priest72a}, shear flow \citep[e.g.,][]{agrimson01a, gavrishchaka99a}, nonlinear whistler wave decay \citep[e.g.,][]{saito17a}, finite electron heat flux \citep[e.g.,][]{dum80a}, nonlinear Langmuir wave decay \citep[e.g.,][]{cairns18a, kellogg13a, zakharov72a, zakharov72b}, and ion-ion instabilities \citep[e.g.,][]{ashourabdalla86a, goodrich18c, goodrich19a, winske87a} have all been shown to reduce or eliminate this temperature ratio threshold.

\indent  Regardless, the occurrence rates of $\tensor*{ \mathcal{T} }{^{ec}_{p}}{\scriptstyle_{j}}$ $\geq$ 3 and $\tensor*{ \mathcal{T} }{^{eff}_{p}}{\scriptstyle_{j}}$ $\geq$ 3 were calculated for reference.  For \textit{Criteria AT}, $\tensor*{ \mathcal{T} }{^{ec}_{p}}{\scriptstyle_{j}}$ $\geq$ 3 is satisfied for $\sim$29.5\%, $\sim$22.6\%, and $\sim$24.4\% of the VDFs for $j$ $=$ $\parallel$, $\perp$, and $tot$, respectively.  These occurrence rates jump to $\sim$34.8\%, $\sim$27.2\%, and $\sim$28.6\% for $\tensor*{ \mathcal{T} }{^{eff}_{p}}{\scriptstyle_{j}}$ $\geq$ 3.  If the data are limited to \textit{Criteria UP}, the occurrence rates for $\tensor*{ \mathcal{T} }{^{ec}_{p}}{\scriptstyle_{j}}$ $\geq$ 3 increase to $\sim$38.2\%, $\sim$38.2\%, and $\sim$37.7\% and similarly the rates for $\tensor*{ \mathcal{T} }{^{eff}_{p}}{\scriptstyle_{j}}$ $\geq$ 3 increase as well to $\sim$47.4\%, $\sim$46.1\%, and $\sim$42.8\%.  If the time range is limited to -120 s $\leq$ $\Delta t$ $\leq$ $+$3 s (where $\Delta t$ is the time from ramp center), the rates for $\tensor*{ \mathcal{T} }{^{eff}_{p}}{\scriptstyle_{j}}$ $\geq$ 3 are $\sim$41.2\%, $\sim$32.3\%, and $\sim$26.4\%.  Thus, even if the analysis is limited to times near the shock ramps the occurrence rates of temperature ratios meeting or exceeding three are smaller than all upstream observations.  The occurrence rate of IAWs has been shown to peak within collisionless shock ramps \citep[e.g.,][]{cohen19a, davis20a, fuselier84a, goodrich18c, wilsoniii07a, wilsoniii14a, wilsoniii14b} and this rate was found to be independent of $\tensor*{ \mathcal{T} }{^{e}_{p}}{\scriptstyle_{tot}}$ \citep[e.g.,][]{wilsoniii07a}.  That the rate of $\tensor*{ \mathcal{T} }{^{eff}_{p}}{\scriptstyle_{j}}$ $\geq$ 3 does not peak near the ramp regions of the 52 IP shocks examined herein is consistent with the apparent lack of dependence on $\tensor*{ \mathcal{T} }{^{e}_{i}}{\scriptstyle_{tot}}$ found in previous work \citep[e.g.,][]{rodriguez75a, wilsoniii07a}.  Note that these rates are significantly higher than those estimated for the ambient solar wind in a recent study \citep[e.g.,][]{wilsoniii18b}.

\indent  The critical drift speed between electrons and ions for a current-driven IAW can be analytically derived for two isotropic Maxwellians that generate a current.  The critical drift is known to depend upon $\tensor*{ \mathcal{T} }{^{e}_{i}}{\scriptstyle_{tot}}$ \citep[e.g.,][]{gurnett05}.  This critical drift threshold for current-driven IAWs, ignoring the details of the ion and electron populations, is satisfied for $\sim$4.78\% of the \totalffpfitsA~VDFs for \textit{Criteria AT} that had solutions for both electron and proton data.  However, as discussed in \citet[][]{priest72a}, the presence of gradients in the temperature and density reduce this idealized critical drift by factors of $\sim$2--8.  This increases the number of VDFs satisfying the critical drift threshold to $\sim$5.2--10.4\%.  These rates require context to appreciate their magnitudes.

\indent  To provide context, some statistical calculations will be performed based upon the observations and relying upon the near ubiquity of IAWs in and around collisionless shock waves \citep[e.g.,][]{breneman13a, chen18a, fuselier84a, goodrich18a, gurnett79a, wilsoniii07a}.  The typical collisionless shock ramp thickness is anywhere from $\gtrsim$1 $\langle \lambda{\scriptstyle_{e}} \rangle{\scriptstyle_{up}}$ to $\lesssim$43 $\langle \lambda{\scriptstyle_{e}} \rangle{\scriptstyle_{up}}$ $\sim$ 1 $\langle \lambda{\scriptstyle_{i}} \rangle{\scriptstyle_{up}}$ \citep[e.g.,][]{hobara10a, mazelle10a}.  For the 52 IP shocks examined herein, the following are satisfied for 90\% of the events:
\begin{itemize}[itemsep=0pt,parsep=0pt,topsep=0pt]
  \item  1.2 km $\lesssim$ $\langle \lambda{\scriptstyle_{e}} \rangle{\scriptstyle_{up}}$ $\lesssim$ 4.2 km;
  \item  155 km/s $\lesssim$ $\langle \lvert V{\scriptstyle_{shn}} \rvert \rangle{\scriptstyle_{up}}$ $\lesssim$ 700 km/s; and
  \item  1 km $\lesssim$ $L{\scriptstyle_{sh}}$ $\lesssim$ 180 km.
\end{itemize}
\noindent  These shock ramp thicknesses correspond to time scales of $\sim$0.002--1.2 seconds in the spacecraft frame, or $\sim$10$^{-5}$--0.008\% of the total time window examined for each IP shock.  The duration of each electron VDF is $\sim$3 seconds\footnote{The instrument is actually triggered on the sun pulse from the sun sensor, which depends upon the spin rate of the spacecraft bus.  The spin period for \emph{Wind} has remained near $\sim$3 seconds for the entirety of the mission but varies by a $\sim$0.1 seconds depending on date and time.}, which means the shock ramps at most $\sim$30\% of the minimum cadence of the \emph{Wind} 3DP instrument.  Thus, any given VDF cannot parameterize only the shock ramp and the instruments cannot directly measure the shock ramp currents.

\indent  Despite the limitation of the particle data time resolution and unrealistic VDF profile assumptions in the theory, there were still nearly 600 VDFs that satisfied the critical drift threshold for current-driven IAWs.  This corresponds to roughly 11 VDFs for each of the 52 IP shocks examined herein or a total duration more than 27 times that of the longest shock ramp in this study.  Further, there were nearly 3500 VDFs (or $\sim$67 per IP shock) that satisfy $\tensor*{ \mathcal{T} }{^{eff}_{p}}{\scriptstyle_{j}}$ $\geq$ 3, i.e., the often quoted temperature ratio threshold for IAWs.  The IAWs of interest here should affect the electron VDFs on time scales much shorter than the integration time, thus the \emph{Wind} 3DP instrument should only observe the post-instability form.

\indent  The \textit{Criteria DN} VDFs have larger core self-similar exponents ($s{\scriptstyle_{ec}}$, $p{\scriptstyle_{ec}}$, and $q{\scriptstyle_{ec}}$) than \textit{Criteria UP} VDFs, with the strongest shocks consistently showing flattops (i.e., $p{\scriptstyle_{ec}}$ $\geq$ 4) for minutes to hours in the downstream, similar to terrestrial bow shock observations.  The larger core exponents in the downstream regions combined with IAW amplitudes positively correlated with Mach number\citep[e.g.,][]{wilsoniii07a} is consistent with IAWs stochastically accelerating the core electrons \citep[e.g.,][]{dum74a, dum75a}.  This type of stochastic acceleration is qualitatively referred to as inelastic scattering throughout this three-part study.  However, such a change in VDF profile has also been interpreted as due to the acceleration by quasi-static cross-shock electric fields \citep[e.g.,][]{hull01a, mitchell14a, schwartz88a, schwartz11a, schwartz14a, scudder86c}.  Note that for both $s{\scriptstyle_{ec}}$ and $p{\scriptstyle_{ec}}$ the differences between \textit{Criteria LM} and \textit{Criteria HM} shocks are not statistically significant.  That is, theory and observation suggest that the magnitude of both the quasi-static cross-shock electric fields and IAW amplitudes should increase with increasing Mach number.  Therefore, if IAWs are affecting the core electrons are they increasing the exponent or the temperature or both?  If so, why and how?  Similar questions arise regarding the quasi-static cross-shock electric field explanation.  Kinetic simulations are required to resolve this discrepancy and are beyond the scope of this study.

\indent  Finally, two more instabilities will be discussed that also affect the ions since most previous instability work has focused on the ions \citep[e.g.,][]{kasper13a, klein17c, klein18a, maruca12a}.  The purpose is to examine the differences in the particle populations near IP shocks versus what is typically considered ambient solar wind.  The threshold for both the mirror and firehose instabilities can also be calculated following the approach\footnote{Note that the major differences between \citet[][]{chen16b} and this work are that: this work does not include secondary proton beams; this work separates the electron components rather than treating them all as one population; and this work includes IP shocks.} in \citet[][]{chen16b}.  Using only VDF solutions when all five particle populations have finite velocity moments, the plasma is unstable to the firehose instability $\sim$1.3\% of the time and mirror $\sim$13.5\%.  These rates are $\sim$10 and $\sim$20 times larger, respectively, than the rates found by \citet[][]{chen16b}.  It may not be surprising to find that VDFs are statistically more unstable near IP shocks than the ambient solar wind, since the shock itself constant is a source of multiple types of free energy.  However, these rates do not significantly change if the time range of analysis is limited to $\pm$20 minutes of the shock ramp center suggesting the enhanced rates are either due to the separation of electron components or the lack of inclusion of a secondary proton beam.  Despite this the \textit{Criteria UP} VDFs should be treated as generally being more unstable than time periods that intentionally avoid IP shocks.

\indent  As a final note, the instability analysis presented in this section should be interpreted with care.  The rates calculated are based upon thresholds and do not directly imply anything about whether an observable wave amplitude would result.  For instance, the threshold for the whistler instabilities are for growth rates at 10\% of the proton cyclotron frequency, i.e., $>$18,000 times longer than a single electron cyclotron period.  Therefore, that only $\sim$3\% of the VDFs are stable does not imply large amplitude whistler waves should be observed for nearly all intervals examined herein\footnote{However, such high rates are consistent with previous studies finding whistler waves to be common downstream of IP shocks \citep[e.g.,][]{coroniti82a, lengyelfrey94, lengyelfrey96} and in the ambient solar wind \citep[e.g.,][]{lengyelfrey96, neubauer77a}.}.  That is, in the cases where the instability thresholds are barely surpassed, the resulting fluctuation amplitudes may be so small that they are below the noise floor of many instruments.  Further, the separation of the electron VDF into three populations, all non-Maxwellian, is different than the two drifting bi-Maxwellians assumed by \citet[][]{gary94a}.  Thus, the large fraction of VDFs satisfying instability thresholds presented herein should not be interpreted as most time intervals exhibiting large amplitude electromagnetic fluctuations.

\phantomsection   
\section{Energy Partition}  \label{sec:EnergyPartition}

\phantomsection   
\subsection{Temperature Differences}  \label{subsec:TemperatureDifferences}

\indent  In this section, the changes in the fit results across all shocks are examined.  The data are presented as the median of all permutations of the difference, $\widetilde{\Delta Q}$, of parameter $Q$ between the upstream and downstream values for each IP shock.  The uncertainties for the fit parameters are half the magnitude of the difference between $X{\scriptstyle_{5\%}}$ and $X{\scriptstyle_{95\%}}$.  The uncertainties for the macroscopic shock parameters result from the standard propagation of uncertainties given by the values in the \emph{Wind} shock database.

\indent  These medians with uncertainties were then fit to a model power-law function, $Y$ $=$ $A \ X^{B}$ $+$ $C$, assuming Poisson weights, where $X$ is one of the macroscopic shock parameters and $Y$ is one of the medians of all permutations of the difference (or ratio) across the shock.  Given that the shock must transform the change in bulk kinetic energy, the obvious shock parameter to examine is the change in kinetic energy, $\Delta \overline{KE}{\scriptstyle_{shn}}$ (see Appendix \ref{app:Definitions}), across the shock\footnote{All other relevant shock parameters were examined and weaker relationships were found between $\langle M{\scriptstyle_{f}} \rangle{\scriptstyle_{up}}$ (not shown) and the permuted differences of $\beta{\scriptstyle_{eb, j}}$ and $P{\scriptstyle_{ec, j}}$.  Weak relationships were also observed between $\Delta \bar{U}{\scriptstyle_{shn}}$ (not shown) and the permuted differences of the following: $\mathcal{A}{\scriptstyle_{ec}}$, $\mathcal{A}{\scriptstyle_{eff}}$, $T{\scriptstyle_{ec, j}}$, $T{\scriptstyle_{eff, j}}$, $T{\scriptstyle_{p, j}}$, and $T{\scriptstyle_{i, j}}$.}.

\begin{figure*}[!htb]
  \centering
    {\includegraphics[trim = 0mm 0mm 0mm 0mm, clip, width=150mm]{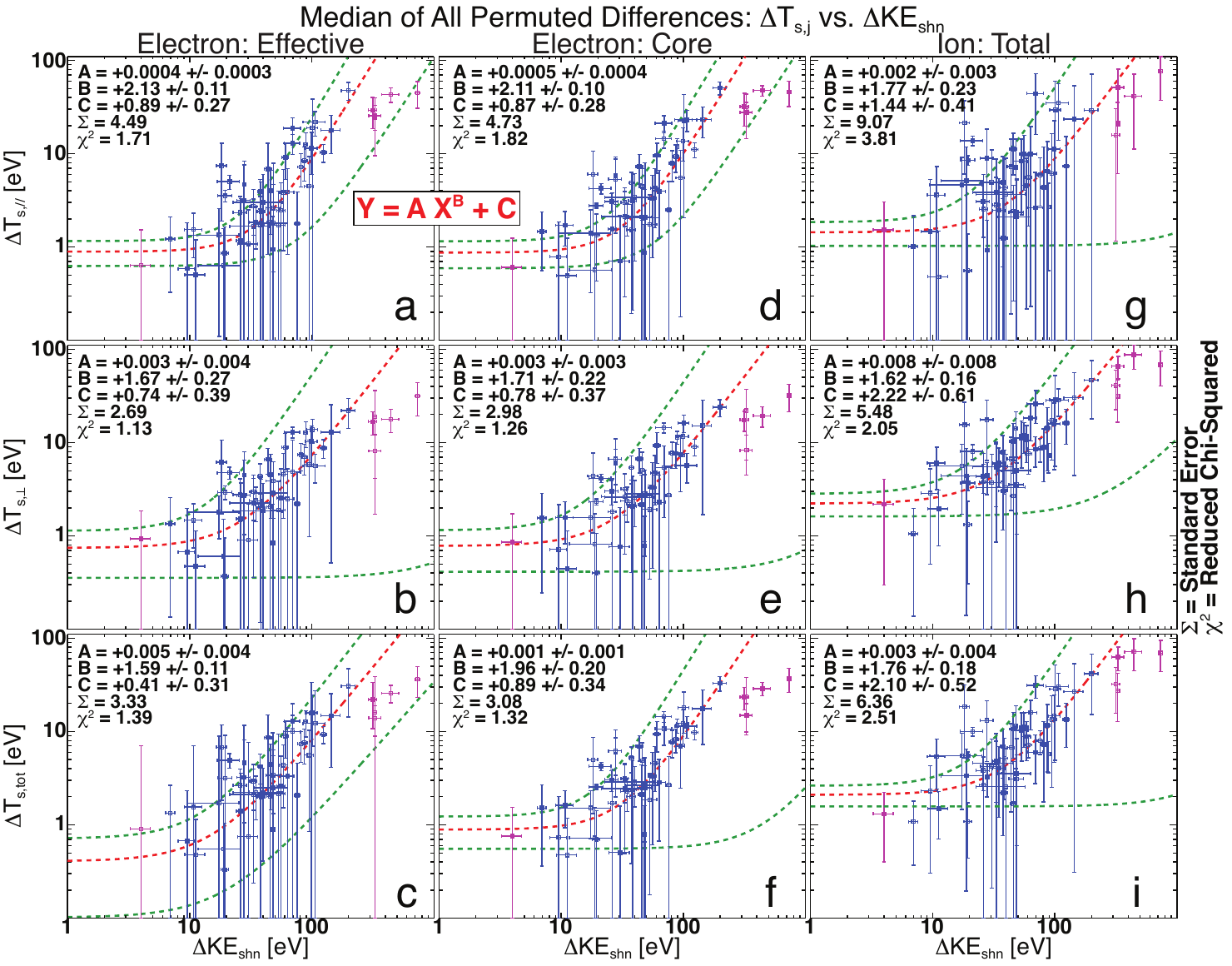}}
    \caption{Median of the permuted temperature differences, $\widetilde{\Delta T}{\scriptstyle_{s, j}}$, versus the change in shock kinetic energy, $\Delta \overline{KE}{\scriptstyle_{shn}}$ [$eV$], for the effective electron temperatures (first column), core electron temperatures (second column), and total ion temperatures (third column).  The first, second, and third rows show the parallel, perpendicular, and total temperature changes, respectively.  The red dashed line in every panel is a power-law fit (function defined between panels a and d) to the data and the green lines show the associated uncertainty bounds.  The associated fit parameters are shown in each panel.  The magenta colored points were ignored during the fit process as outliers.}
    \label{fig:TempDifferences}
\end{figure*}

\indent  Figure \ref{fig:TempDifferences} shows the relationship between the change in temperature versus the change in shock kinetic energy and Table \ref{tab:FitParamsDiffs} shows the values of the fit parameters with goodness of fit estimates.  Here, the $\widetilde{\Delta T}{\scriptstyle_{s, j}}$ values represent the median\footnote{Note that the use of the average of all permutations of the differences did not yield significantly different results than the median.} of all the permuted temperature differences between downstream and upstream for each IP shock.  The $\Delta \overline{KE}{\scriptstyle_{shn}}$ values are just computed from the $\langle \lvert U{\scriptstyle_{shn}} \rvert \rangle{\scriptstyle_{j}}$ values from the \emph{Wind} shock database.  The uncertainties are calculated using the standard propagation of uncertainties.

\startlongtable  
\begin{deluxetable}{| l | c | c | c | c | c |}
  \tabletypesize{\footnotesize}    
  \tablecaption{Fit Parameters for Figure \ref{fig:TempDifferences} \label{tab:FitParamsDiffs}}
  \tablehead{\colhead{Y} & \colhead{A $\times 10^{-3}$}\tablenotemark{a} & \colhead{B}\tablenotemark{b} & \colhead{C}\tablenotemark{c} & \colhead{$\tilde{\chi}^{2}$}\tablenotemark{d} & \colhead{$\Sigma{\scriptstyle_{fit}}$}\tablenotemark{e}}
  \startdata
  $T{\scriptstyle_{eff, \parallel}}$ & 0.4 $\pm$ 0.3\tablenotemark{f} & 2.13 $\pm$ 0.11 & 0.89 $\pm$ 0.27 & 1.7 & 4.5  \\
  $T{\scriptstyle_{eff, \perp}}$     & 0.3 $\pm$ 0.4 & 1.67 $\pm$ 0.27 & 0.74 $\pm$ 0.39 & 1.1 & 2.7  \\
  $T{\scriptstyle_{eff, tot}}$       & 5.0 $\pm$ 4.0 & 1.59 $\pm$ 0.11 & 0.41 $\pm$ 0.31 & 1.4 & 3.3  \\
  $T{\scriptstyle_{ec, \parallel}}$  & 0.5 $\pm$ 0.5 & 2.11 $\pm$ 0.10 & 0.87 $\pm$ 0.28 & 1.8 & 4.7  \\
  $T{\scriptstyle_{ec, \perp}}$      & 3.0 $\pm$ 3.0 & 1.71 $\pm$ 0.22 & 0.78 $\pm$ 0.37 & 1.3 & 3.0  \\
  $T{\scriptstyle_{ec, tot}}$        & 1.0 $\pm$ 1.0 & 1.96 $\pm$ 0.20 & 0.89 $\pm$ 0.34 & 1.3 & 3.1  \\
  $T{\scriptstyle_{i, \parallel}}$   & 2.0 $\pm$ 3.0 & 1.77 $\pm$ 0.23 & 1.44 $\pm$ 0.41 & 3.8 & 9.1  \\
  $T{\scriptstyle_{i, \perp}}$       & 8.0 $\pm$ 8.0 & 1.61 $\pm$ 0.16 & 2.22 $\pm$ 0.60 & 2.1 & 5.5  \\
  $T{\scriptstyle_{i, tot}}$         & 3.0 $\pm$ 4.0 & 1.76 $\pm$ 0.18 & 2.10 $\pm$ 0.52 & 2.5 & 6.4  \\
  \hline
  \enddata
  \tablenotetext{a}{constant multiplier in power-law}
  \tablenotetext{b}{exponent in power-law}
  \tablenotetext{c}{constant offset in power-law}
  \tablenotetext{d}{reduced chi-squared of fit}
  \tablenotetext{e}{standard error between fit and data}
  \tablenotetext{f}{values shown are larger by a factor of $10^{+3}$}
  \tablecomments{For symbol definitions, see Appendix \ref{app:Definitions}.}
\end{deluxetable}

\indent  The first thing to notice is that the relationship is not linear.  In fact, several previous studies examined the change in temperature and found positive correlations between $\Delta \bar{T}{\scriptstyle_{e, tot}}$ and variants of $\Delta \overline{KE}{\scriptstyle_{shn}}$ \citep[e.g.,][]{hull00a, schwartz88a, thomsen87b, thomsen93a}.  However, if one examines the plots of $\widetilde{\Delta T}{\scriptstyle_{e, tot}}$ versus $\Delta \overline{KE}{\scriptstyle_{shn}}$ in each of these studies, it's not clear whether the trend is linear or otherwise.  For instance, examination of Figures 5 and 6 in \citet[][]{feldman83c} do not appear to have a linear trend.  The linear relationship between $\Delta \bar{T}{\scriptstyle_{e, tot}}$ and $\Delta \overline{KE}{\scriptstyle_{shn}}$ found by \citet[][]{hull00a} required that they ignore shocks with $\Delta \overline{KE}{\scriptstyle_{shn}}$ $<$ 100 eV in the fit.  The relationship between $\Delta \bar{T}{\scriptstyle_{e, tot}}$ and $\Delta \bar{U}{\scriptstyle_{shn}}^{2}$ found by \citet[][]{fitzenreiter03a}, however, did appear to be linear.  Note that many of the previous studies examined the higher Mach number terrestrial bow shock.

\indent  However, it is not unreasonable to assume that a linear relationship would exist if the energy conversion was a linear.  That is, if the irreversible transformation of excess kinetic energy across the shock\footnote{Some of the total energy transformation must be irreversible to initiate a shock from a nonlinearly steepening wave but once initiated, reversible processes can maintain the shock \citep[e.g.,][]{shu92a}.} into particle heating occurred directly with no intermediary processes\footnote{i.e., $\Delta \overline{KE}{\scriptstyle_{shn}}$ $\rightarrow$ $\Delta \bar{T}{\scriptstyle_{e, tot}}$}, one may expect that a linear relationship would exist with the change in electron and ion temperature with some associated efficiency.  A linear relationship also requires, in general, fewer assumptions to model.  That is, the scatter plot of points in previous studies could have just as easily been fit to a nonlinear function like a power-law though it is likely these authors chose a linear relationship as that is the simplest possibility.

\indent  One should also note that the exponents for $\widetilde{\Delta T}{\scriptstyle_{ec, \parallel}}$ and $\widetilde{\Delta T}{\scriptstyle_{eff, \parallel}}$ are larger than all exponents of $\widetilde{\Delta T}{\scriptstyle_{i, j}}$.  However, if the five magenta points at large $\Delta \overline{KE}{\scriptstyle_{shn}}$ are included in the fits, the exponents for $\widetilde{\Delta T}{\scriptstyle_{ec, j}}$ and $\widetilde{\Delta T}{\scriptstyle_{eff, j}}$ would decrease more than $\widetilde{\Delta T}{\scriptstyle_{i, j}}$.  Note that previous work has either inferred \citep[e.g.,][]{ghavamian07a, ghavamian14a} or showed with in situ measurements that stronger shocks heat ions more than electrons \citep[e.g.,][]{schwartz88a, thomsen93a, masters11a}.  In fact, \citet[][]{schwartz88a} and others have even noticed that $\Delta \bar{T}{\scriptstyle_{e, tot}}$/$\Delta \bar{T}{\scriptstyle_{t, tot}}$ $\propto$ $\langle M{\scriptstyle_{A}} \rangle{\scriptstyle_{up}}^{-1}$.  Therefore, the change in slope/trend at larger $\Delta \overline{KE}{\scriptstyle_{shn}}$ may be indicative of differences in shock energy dissipation at stronger shocks, as suggested by trends in previous work.

\indent  Even when the five magenta points at large $\Delta \overline{KE}{\scriptstyle_{shn}}$ (i.e., magenta points in upper right-hand corner of each panel of Figure \ref{fig:TempDifferences}) were included in the fit, the linear fit had much larger $\tilde{\chi}^{2}$ or $\Sigma{\scriptstyle_{fit}}$ values (not shown).  Though it is also fair to argue that the fits shown in Figure \ref{fig:TempDifferences} are not really good fits despite the low $\tilde{\chi}^{2}$ or $\Sigma{\scriptstyle_{fit}}$ values shown in Table \ref{tab:FitParamsDiffs}.  That is, the data have large uncertainties and large relative spread for each $\Delta \overline{KE}{\scriptstyle_{shn}}$, as evidenced by the green lines on either side of the red fit lines.  Therefore, it is likely that if even stronger shocks were added to this data set the power-law trend presented in Figure \ref{fig:TempDifferences} would need to be modified by either an exponential roll-over or higher order terms or the trend would entirely fall apart.  Thus, these fits should be treated with caution and/or skepticism.

\indent  To verify that the use of the median on the permutations of all differences was not causing the nonlinearity, the same temperature differences were plotted versus $\Delta \overline{KE}{\scriptstyle_{shn}}$ but now using $\Delta \bar{T}{\scriptstyle_{s, j}}$ instead of $\widetilde{\Delta T}{\scriptstyle_{s, j}}$ (not shown).  That is, the average over each region was calculated as a single scalar prior to finding the difference between the regions.  The nonlinear fit lines were still a better match to the results than a linear line.  Thus, the nonlinear relationship between the temperature increase and the change in kinetic energy across the shock appears to be real at least for low $\Delta \overline{KE}{\scriptstyle_{shn}}$.

\indent  Most of the earlier work looking at the dependence of $\Delta \bar{T}{\scriptstyle_{s, j}}$ on $\Delta \overline{KE}{\scriptstyle_{shn}}$ focused on the Earth's bow shock, which is typically higher Mach number (i.e., $\langle M{\scriptstyle_{f}} \rangle{\scriptstyle_{up}}$ $\gtrsim$ 3--10) than those examined herein.  The Mach numbers in this study satisfy 1.01 $\leq$ $\langle M{\scriptstyle_{f}} \rangle{\scriptstyle_{up}}$ $\leq$ 6.4, with 90\% satisfying $\sim$1.15--4.00, and a median of $\sim$1.86.  Note that 45 of the 52 IP shocks examined herein satisfied $\langle M{\scriptstyle_{f}} \rangle{\scriptstyle_{up}}$ $<$ 3.  Further, the relationship between $\langle M{\scriptstyle_{f}} \rangle{\scriptstyle_{up}}$ and $\Delta \overline{KE}{\scriptstyle_{shn}}$ is not linear.  This begs the question of what could cause the energy transformation from $\Delta \overline{KE}{\scriptstyle_{shn}}$ into $\Delta \bar{T}{\scriptstyle_{s, j}}$ to be nonlinear, if the trend is real.

\indent  Regardless, the fraction of $\Delta \overline{KE}{\scriptstyle_{shn}}$ distributed to each of the populations shown in Figure \ref{fig:TempDifferences}, for 90\% of events (i.e., $X{\scriptstyle_{5\%}}$ to $X{\scriptstyle_{95\%}}$ range), satisfy the following:
\begin{itemize}[itemsep=0pt,parsep=0pt,topsep=0pt]
  \item  1.5\% $\lesssim$ $\widetilde{\Delta T}{\scriptstyle_{ec, j}} / \Delta \overline{KE}{\scriptstyle_{shn}}$ $\lesssim$ 34\%;
  \item  0.8\% $\lesssim$ $\widetilde{\Delta T}{\scriptstyle_{eff, j}} / \Delta \overline{KE}{\scriptstyle_{shn}}$ $\lesssim$ 41\%; and
  \item  1.4\% $\lesssim$ $\widetilde{\Delta T}{\scriptstyle_{i, j}} / \Delta \overline{KE}{\scriptstyle_{shn}}$ $\lesssim$ 63\%.
\end{itemize}
\noindent  Although these ratios are mostly uniform for all $j$ for both electron populations, the $\widetilde{\Delta T}{\scriptstyle_{i, \parallel}} / \Delta \overline{KE}{\scriptstyle_{shn}}$ range is systematically smaller than the other two components.  That is, if one separates the perpendicular and total from the parallel components then 90\% of the events would satisfy:
\begin{itemize}[itemsep=0pt,parsep=0pt,topsep=0pt]
  \item  5.8\% $\lesssim$ $\widetilde{\Delta T}{\scriptstyle_{i, \{\perp, tot\} }} / \Delta \overline{KE}{\scriptstyle_{shn}}$ $\lesssim$ 63\%; and
  \item  1.4\% $\lesssim$ $\widetilde{\Delta T}{\scriptstyle_{i, \parallel}} / \Delta \overline{KE}{\scriptstyle_{shn}}$ $\lesssim$ 40\%.
\end{itemize}
\noindent  Similarly the median values for $\widetilde{\Delta T}{\scriptstyle_{ec, j}} / \Delta \overline{KE}{\scriptstyle_{shn}}$ and $\widetilde{\Delta T}{\scriptstyle_{eff, j}} / \Delta \overline{KE}{\scriptstyle_{shn}}$ are in the $\sim$7.6--9.3\% range while the $\widetilde{\Delta T}{\scriptstyle_{i, j}} / \Delta \overline{KE}{\scriptstyle_{shn}}$ median values satisfy $\sim$10.7--17.9\%, with the parallel component being the smallest.  In summary, the electron core and effective populations only gain $\sim$40--80\% of what the ions do in thermal energy across the shock for these events.

\phantomsection   
\subsection{Energy Density Differences}  \label{subsec:EnergyDensityDifferences}

\indent  In this section, the partition of energy among the five primary constituent particle populations will be discussed.  These five are the electron core ($s$ $=$ $ec$), halo ($s$ $=$ $eh$), beam/strahl ($s$ $=$ $eb$), and ion proton ($s$ $=$ $p$) and alpha-particle ($s$ $=$ $\alpha$) populations.  Similar to Section \ref{subsec:TemperatureDifferences}, the median of all permutations of the differences will be discussed.

\indent  The normalized pressures, $\widetilde{\Delta \Pi}{\scriptstyle_{s, j}}$ and $\widetilde{\Delta \psi}{\scriptstyle_{s, j}}$ were plotted (not shown) versus $\theta{\scriptstyle_{Bn}}$, $\langle \lvert V{\scriptstyle_{shn}} \rvert \rangle{\scriptstyle_{up}}$, $\langle \lvert U{\scriptstyle_{shn}} \rvert \rangle{\scriptstyle_{up}}$, $\langle M{\scriptstyle_{f}} \rangle{\scriptstyle_{up}}$, $\langle M{\scriptstyle_{A}} \rangle{\scriptstyle_{up}}$, $\langle M{\scriptstyle_{Te}} \rangle{\scriptstyle_{up}}$, $\Delta \bar{U}{\scriptstyle_{shn}}$, $\Delta \overline{KE}{\scriptstyle_{shn}}$, and $\Delta \bar{\xi}{\scriptstyle_{shn}}$ (i.e., every macroscopic shock parameter predicted to be of importance here).  In the following, weak correlations imply there is a trend but the large scatter and multiple outliers make interpretation difficult.  Moderate correlations are for clear trends but still with a significant spread in data (e.g., similar to $\widetilde{\Delta T}{\scriptstyle_{s, j}}$ plots in Figure \ref{fig:TempDifferences}).  There were no good correlations observed for any pair of parameters examined, only a few moderate correlations, and several weak correlations (not shown).

\indent  For reference, the following will show parameters as $X{\scriptstyle_{5\%}}$ $\lesssim$ $X$ $\lesssim$ $X{\scriptstyle_{95\%}}$, $\tilde{X}$, for the 52 IP shocks examined herein:
\begin{itemize}[itemsep=0pt,parsep=0pt,topsep=0pt]
  \item  4.07\% $\lesssim$ $\widetilde{\Delta \Pi}{\scriptstyle_{ec, j}}$ $\lesssim$ 41.0\%, $\sim$12.7\%;
  \item  1.15\% $\lesssim$ $\widetilde{\Delta \Pi}{\scriptstyle_{eh, j}}$ $\lesssim$ 10.6\%, $\sim$3.85\%;
  \item  0.30\% $\lesssim$ $\widetilde{\Delta \Pi}{\scriptstyle_{eb, j}}$ $\lesssim$ 7.46\%, $\sim$3.22\%;
  \item  4.36\% $\lesssim$ $\widetilde{\Delta \Pi}{\scriptstyle_{eff, j}}$ $\lesssim$ 36.1\%, $\sim$12.8\%;
  \item  4.05\% $\lesssim$ $\widetilde{\Delta \Pi}{\scriptstyle_{p, j}}$ $\lesssim$ 36.1\% , $\sim$12.5\%; and
  \item  0.15\% $\lesssim$ $\widetilde{\Delta \Pi}{\scriptstyle_{\alpha, j}}$ $\lesssim$ 9.05\%, $\sim$2.33\%.
\end{itemize}
\noindent  If the analysis is limited to \textit{Criteria LM} shocks, then these relations go to:
\begin{itemize}[itemsep=0pt,parsep=0pt,topsep=0pt]
  \item  2.36\% $\lesssim$ $\widetilde{\Delta \Pi}{\scriptstyle_{ec, j}}$ $\lesssim$ 38.3\%, $\sim$12.8\%;
  \item  1.06\% $\lesssim$ $\widetilde{\Delta \Pi}{\scriptstyle_{eh, j}}$ $\lesssim$ 9.49\%, $\sim$3.57\%;
  \item  0.74\% $\lesssim$ $\widetilde{\Delta \Pi}{\scriptstyle_{eb, j}}$ $\lesssim$ 6.87\%, $\sim$2.43\%;
  \item  4.36\% $\lesssim$ $\widetilde{\Delta \Pi}{\scriptstyle_{eff, j}}$ $\lesssim$ 35.3\%, $\sim$12.4\%;
  \item  4.36\% $\lesssim$ $\widetilde{\Delta \Pi}{\scriptstyle_{p, j}}$ $\lesssim$ 35.3\%, $\sim$12.4\%; and
  \item  0.15\% $\lesssim$ $\widetilde{\Delta \Pi}{\scriptstyle_{\alpha, j}}$ $\lesssim$ 9.05\%, $\sim$2.35\%.
\end{itemize}
\noindent  Finally, if the analysis is limited to \textit{Criteria HM} shocks, then these relations go to:
\begin{itemize}[itemsep=0pt,parsep=0pt,topsep=0pt]
  \item  4.51\% $\lesssim$ $\widetilde{\Delta \Pi}{\scriptstyle_{ec, j}}$ $\lesssim$ 43.8\%, $\sim$10.3\%;
  \item  1.83\% $\lesssim$ $\widetilde{\Delta \Pi}{\scriptstyle_{eh, j}}$ $\lesssim$ 11.6\%, $\sim$6.21\%;
  \item  0.97\% $\lesssim$ $\widetilde{\Delta \Pi}{\scriptstyle_{eb, j}}$ $\lesssim$ 7.46\%, $\sim$5.96\%;
  \item  4.40\% $\lesssim$ $\widetilde{\Delta \Pi}{\scriptstyle_{eff, j}}$ $\lesssim$ 40.4\%, $\sim$13.4\%;
  \item  3.73\% $\lesssim$ $\widetilde{\Delta \Pi}{\scriptstyle_{p, j}}$ $\lesssim$ 40.4\%, $\sim$13.7\%; and
  \item  0.43\% $\lesssim$ $\widetilde{\Delta \Pi}{\scriptstyle_{\alpha, j}}$ $\lesssim$ 3.48\%, $\sim$1.68\%.
\end{itemize}

\indent  Note that the \emph{Wind} SWE Faraday cups have difficulty separating alpha-particles from protons and finding good nonlinear fit solutions in the immediate downstream of the stronger shocks in this study, which is likely affecting the upper bounds of both $\widetilde{\Delta \Pi}{\scriptstyle_{p, j}}$ and $\widetilde{\Delta \Pi}{\scriptstyle_{\alpha, j}}$.  The reason for emphasizing this point is that previous work \citep[e.g.,][]{schwartz88a, thomsen93a, masters11a} found more energy going to the ions as the Mach number increases but the core electrons seem comparable to the protons here.  A possible difference may be that previous work examined the total ion and electron VDFs, which may blur the differences in trends between the various components of each species.  It is also worth noting that the $\widetilde{\Delta \Pi}{\scriptstyle_{s, j}}$ values result from the absolute value of the differences, i.e., the actual change may be negative for one population.

\indent  To determine which population gained more thermal energy density across the shocks, the ratio of the electron component to proton thermal energy densities are examined where the the parameters are shown as $X{\scriptstyle_{5\%}}$ $\lesssim$ $X$ $\lesssim$ $X{\scriptstyle_{95\%}}$, $\tilde{X}$.  The values the 52 IP shocks examined herein are as follows:
\begin{itemize}[itemsep=0pt,parsep=0pt,topsep=0pt]
  \item  16.9\% $\lesssim$ $\widetilde{\Delta P}{\scriptstyle_{ec, j}}$/$\widetilde{\Delta P}{\scriptstyle_{p, j}}$ $\lesssim$ 391\%, $\sim$101\%;
  \item  1.11\% $\lesssim$ $\widetilde{\Delta P}{\scriptstyle_{eh, j}}$/$\widetilde{\Delta P}{\scriptstyle_{p, j}}$ $\lesssim$ 41.3\%, $\sim$5.41\%;
  \item  0.47\% $\lesssim$ $\widetilde{\Delta P}{\scriptstyle_{eb, j}}$/$\widetilde{\Delta P}{\scriptstyle_{p, j}}$ $\lesssim$ 18.3\%, $\sim$2.55\%; and
  \item  19.6\% $\lesssim$ $\widetilde{\Delta P}{\scriptstyle_{eff, j}}$/$\widetilde{\Delta P}{\scriptstyle_{p, j}}$ $\lesssim$ 453\%, $\sim$107\%.
\end{itemize}
\noindent  If the analysis is limited to \textit{Criteria LM} shocks, then these relations go to:
\begin{itemize}[itemsep=0pt,parsep=0pt,topsep=0pt]
  \item  15.8\% $\lesssim$ $\widetilde{\Delta P}{\scriptstyle_{ec, j}}$/$\widetilde{\Delta P}{\scriptstyle_{p, j}}$ $\lesssim$ 391\%, $\sim$89.7\%;
  \item  0.94\% $\lesssim$ $\widetilde{\Delta P}{\scriptstyle_{eh, j}}$/$\widetilde{\Delta P}{\scriptstyle_{p, j}}$ $\lesssim$ 28.8\%, $\sim$5.41\%;
  \item  0.47\% $\lesssim$ $\widetilde{\Delta P}{\scriptstyle_{eb, j}}$/$\widetilde{\Delta P}{\scriptstyle_{p, j}}$ $\lesssim$ 22.2\%, $\sim$2.37\%; and
  \item  17.3\% $\lesssim$ $\widetilde{\Delta P}{\scriptstyle_{eff, j}}$/$\widetilde{\Delta P}{\scriptstyle_{p, j}}$ $\lesssim$ 453\%, $\sim$102\%.
\end{itemize}
\noindent  Finally, if the analysis is limited to \textit{Criteria HM} shocks, then these relations go to:
\begin{itemize}[itemsep=0pt,parsep=0pt,topsep=0pt]
  \item  36.4\% $\lesssim$ $\widetilde{\Delta P}{\scriptstyle_{ec, j}}$/$\widetilde{\Delta P}{\scriptstyle_{p, j}}$ $\lesssim$ 372\%, $\sim$182\%;
  \item  2.82\% $\lesssim$ $\widetilde{\Delta P}{\scriptstyle_{eh, j}}$/$\widetilde{\Delta P}{\scriptstyle_{p, j}}$ $\lesssim$ 49.3\%, $\sim$8.76\%;
  \item  0.54\% $\lesssim$ $\widetilde{\Delta P}{\scriptstyle_{eb, j}}$/$\widetilde{\Delta P}{\scriptstyle_{p, j}}$ $\lesssim$ 10.0\%, $\sim$3.42\%; and
  \item  22.0\% $\lesssim$ $\widetilde{\Delta P}{\scriptstyle_{eff, j}}$/$\widetilde{\Delta P}{\scriptstyle_{p, j}}$ $\lesssim$ 391\%, $\sim$199\%.
\end{itemize}
\noindent  Therefore, the change in core electron thermal pressure is comparable to or larger than that for the protons even for the \textit{Criteria HM} shocks, unexpectedly.  For the \textit{Criteria LM} shocks the core electrons receive roughly the same heating as the protons, consistent with previous results \citep[e.g.,][]{schwartz88a}.  Note that despite the use of pressure here as a measure of thermal energy density, none of these processes are occurring in a thermodynamic system.  That is, there is no well defined equation of state for each of the particle populations.

\indent  In summary, none of the energy density ratios showed a clear dependence upon any macroscopic shock parameter.  None of the $\widetilde{\Delta \Pi}{\scriptstyle_{s, j}}$ or $\widetilde{\Delta \psi}{\scriptstyle_{s, j}}$ cared about $\theta{\scriptstyle_{Bn}}$, i.e., the change in the ratio of thermal energy density to both the total pressure and total energy density is independent of shock geometry.  In other words, the shock geometry does not appear to affect the change in the partition of energy amongst the five major particle populations.  The most correlations were found with $\Delta \bar{U}{\scriptstyle_{shn}}$, but again none of them were good.  The only good correlation was observed between $\Delta \bar{\xi}{\scriptstyle_{shn}}$ and $\widetilde{\Delta \epsilon}{\scriptstyle_{j}}$ (not shown), i.e., the change in total energy density increases with increasing change in shock kinetic energy density.  This merely shows that as the available free energy increases, the total internal energy increases, as expected.

\indent  Finally, the absolute changes in normalized partial pressures were dominated by the core electrons and protons with the suprathermal electrons and alpha-particles serving as minor constituents which is again expected.  However, the absolute differences discussed in this section do not inform us whether the change is positive or negative.  Further, somewhat unexpectedly the core electron pressure changes were often larger than that for the protons for \textit{Criteria HM} shocks, while for \textit{Criteria LM} shocks the changes were closer to previous observations \citep[e.g.,][]{schwartz88a}.  Again, some of this is most likely due to the issues facing the \emph{Wind} SWE Faraday cups downstream of the strongest shocks in this study.  However, the larger than unity ratios of $\widetilde{\Delta P}{\scriptstyle_{ec, j}}$/$\widetilde{\Delta P}{\scriptstyle_{p, j}}$ for even \textit{Criteria LM} shocks were not really expected.  To help address this, SEA plots are shown to illustrate the trends in $\Delta \Pi{\scriptstyle_{s, j}}$ versus time in the following section.

\phantomsection   
\subsection{Thermal Energy Density Trends}  \label{subsec:ThermalEnergyDensity}

\indent  In this section, the partition of energy among the five primary constituent particle populations will be discussed.  These five are the electron core ($s$ $=$ $ec$), halo ($s$ $=$ $eh$), beam/strahl ($s$ $=$ $eb$), and ion proton ($s$ $=$ $p$) and alpha-particle ($s$ $=$ $\alpha$) populations.  Similar to Section \ref{sec:SuperposedEpochAnalysis}, SEA plots will be presented.

\indent  Since none of the $\widetilde{\Delta Q}$ seemed to show very good correlations with any macroscopic shock parameter predicted to be of importance, the statistical trend of the normalized energy densities directly using SEA were examined.  The purpose is to see if any clear trend versus time(space) emerges that is not reflected in macroscopic differences or ratios.  Note that the SEA plots involve the calculation of one-variable statistics for all points within a given time bin, while the $\widetilde{\Delta Q}$ values are calculated on a shock-by-shock basis.

\begin{figure}
  \centering
    {\includegraphics[trim = 0mm 0mm 0mm 0mm, clip, width=80mm]{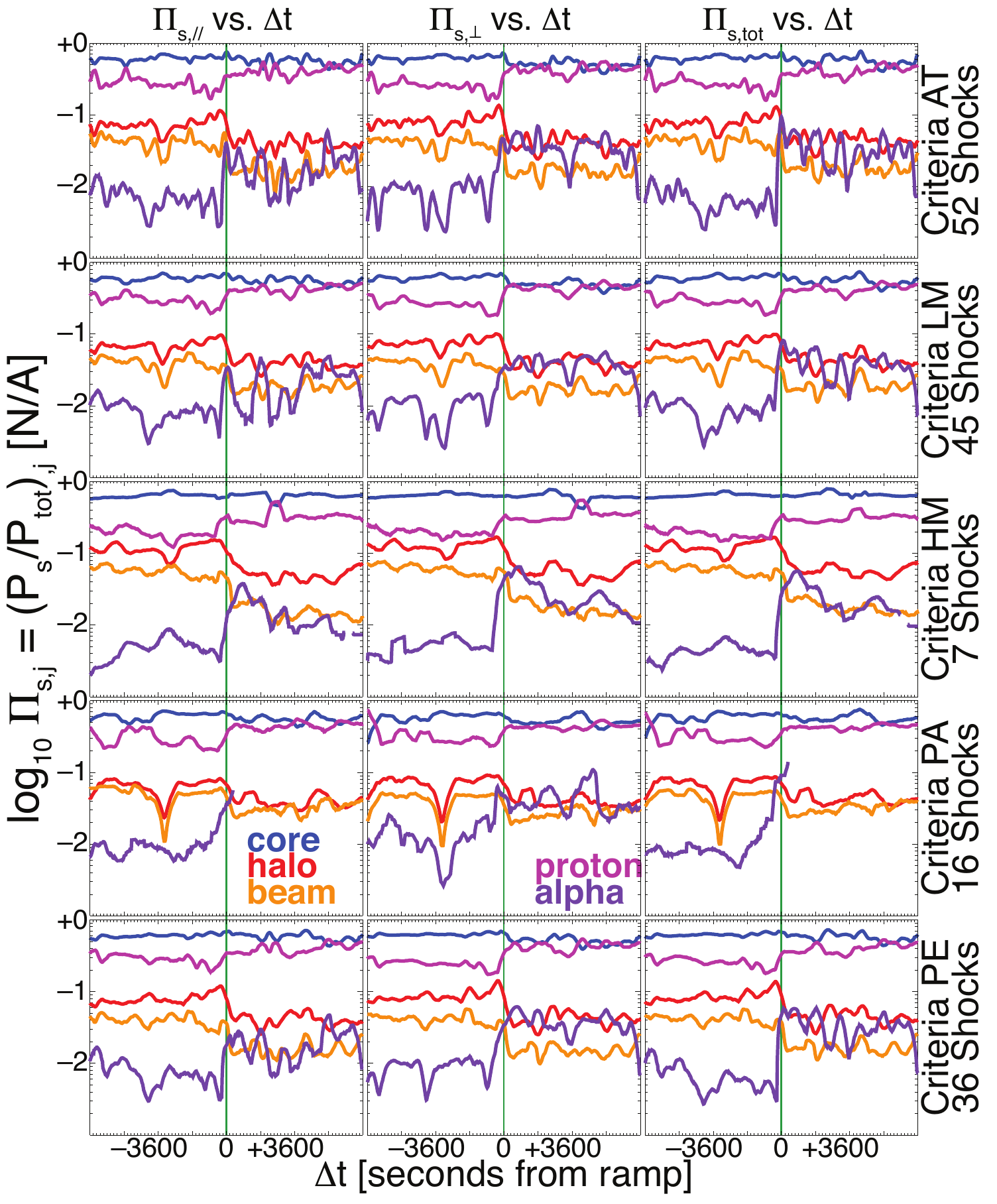}}
    \caption{The running median (e.g., red lines in Figure \ref{fig:Exponents}) from the SEA of the partial thermal pressures, $\Pi{\scriptstyle_{s, j}}$ (see Appendix \ref{app:Definitions} for parameter definitions), for each of the following particle populations:  core electrons (blue lines), halo electrons (red lines), beam/strahl electrons (orange lines), protons (magenta lines), and alpha-particles (purple/violet lines).  Unlike previous SEA plots, these are not normalized to an upstream median value.  The vertical green line indicates the ramp center time.  All panels have uniform horizontal and vertical axis ranges.}
    \label{fig:SEAMedianOnly}
\end{figure}

\indent  Figure \ref{fig:SEAMedianOnly} shows the running median only from SEA of the partial thermal pressures, $\Pi{\scriptstyle_{s, j}}$, of each of the major particle populations in the solar wind, including the core electrons (blue lines), halo electrons (red lines), beam/strahl electrons (orange lines), protons (magenta lines), and alpha-particles (purple/violet lines).  First, $\Pi{\scriptstyle_{ec, j}}$ and $\Pi{\scriptstyle_{p, j}}$ dominate at all times for all selection criteria, as expected.  Second, the general trend of all electron populations is for their fractional thermal energy density to reduce across the shock ramp except for $\Pi{\scriptstyle_{ec, j}}$ for \textit{Criteria HM} shocks.  Both the $\Pi{\scriptstyle_{eh, j}}$ and $\Pi{\scriptstyle_{eb, j}}$ electrons consistently decrease across the shock, with the weakest change across \textit{Criteria PA} shocks.  The $\Pi{\scriptstyle_{eb, j}}$ show a continual decreasing trend across \textit{Criteria HM} shocks with a distinct jump at the shock ramp in contrast to $\Pi{\scriptstyle_{eh, j}}$ which basically levels off and recovers downstream.  There are also several intervals where $\Pi{\scriptstyle_{ec, j}}$ and $\Pi{\scriptstyle_{p, j}}$ seem to oscillate exactly out of phase from each other, likely owing to pressure balance features near these shocks.  The interesting aspect is that they are common/strong enough to show up in a running median constructed from SEA on 52 different shocks.

\indent  In general, the following are satisfied for $\sim$90\% of all data for \textit{Criteria AT} (i.e., between $X{\scriptstyle_{5\%}}$ and $X{\scriptstyle_{95\%}}$), from largest to smallest (see Table \ref{tab:PressureRatios}):
\begin{itemize}[itemsep=0pt,parsep=0pt,topsep=0pt]
  \item  25\% $\lesssim$ $\Pi{\scriptstyle_{ec, j}}$ $\lesssim$ 92\%;
  \item  13\% $\lesssim$ $\Pi{\scriptstyle_{p, j}}$ $\lesssim$ 72\%;
  \item  1\% $\lesssim$ $\Pi{\scriptstyle_{eh, j}}$ $\lesssim$ 18\%;
  \item  0.3\% $\lesssim$ $\Pi{\scriptstyle_{eb, j}}$ $\lesssim$ 11\%; and
  \item  0.2\% $\lesssim$ $\Pi{\scriptstyle_{\alpha, j}}$ $\lesssim$ 11\%.
\end{itemize}
\noindent  These are the partitions of thermal energy density for all time periods.  When the data are separated into upstream and downstream, things change slightly but not tremendously.  For \textit{Criteria UP}, sorted from largest to smallest, the following are satisfied:
\begin{itemize}[itemsep=0pt,parsep=0pt,topsep=0pt]
  \item  27\% $\lesssim$ $\Pi{\scriptstyle_{ec, j}}$ $\lesssim$ 85\%;
  \item  11\% $\lesssim$ $\Pi{\scriptstyle_{p, j}}$ $\lesssim$ 68\%;
  \item  1\% $\lesssim$ $\Pi{\scriptstyle_{eh, j}}$ $\lesssim$ 23\%;
  \item  0.7\% $\lesssim$ $\Pi{\scriptstyle_{eb, j}}$ $\lesssim$ 15\%; and
  \item  0.1\% $\lesssim$ $\Pi{\scriptstyle_{\alpha, j}}$ $\lesssim$ 9\%.
\end{itemize}
\noindent  For \textit{Criteria DN}, sorted from largest to smallest, the following are satisfied:
\begin{itemize}[itemsep=0pt,parsep=0pt,topsep=0pt]
  \item  24\% $\lesssim$ $\Pi{\scriptstyle_{ec, j}}$ $\lesssim$ 95\%;
  \item  15\% $\lesssim$ $\Pi{\scriptstyle_{p, j}}$ $\lesssim$ 74\%;
  \item  1\% $\lesssim$ $\Pi{\scriptstyle_{eh, j}}$ $\lesssim$ 15\%;
  \item  0.2\% $\lesssim$ $\Pi{\scriptstyle_{eb, j}}$ $\lesssim$ 8\%; and
  \item  0.2\% $\lesssim$ $\Pi{\scriptstyle_{\alpha, j}}$ $\lesssim$ 13\%.
\end{itemize}

\begin{figure}
  \centering
    {\includegraphics[trim = 0mm 0mm 0mm 0mm, clip, width=80mm]{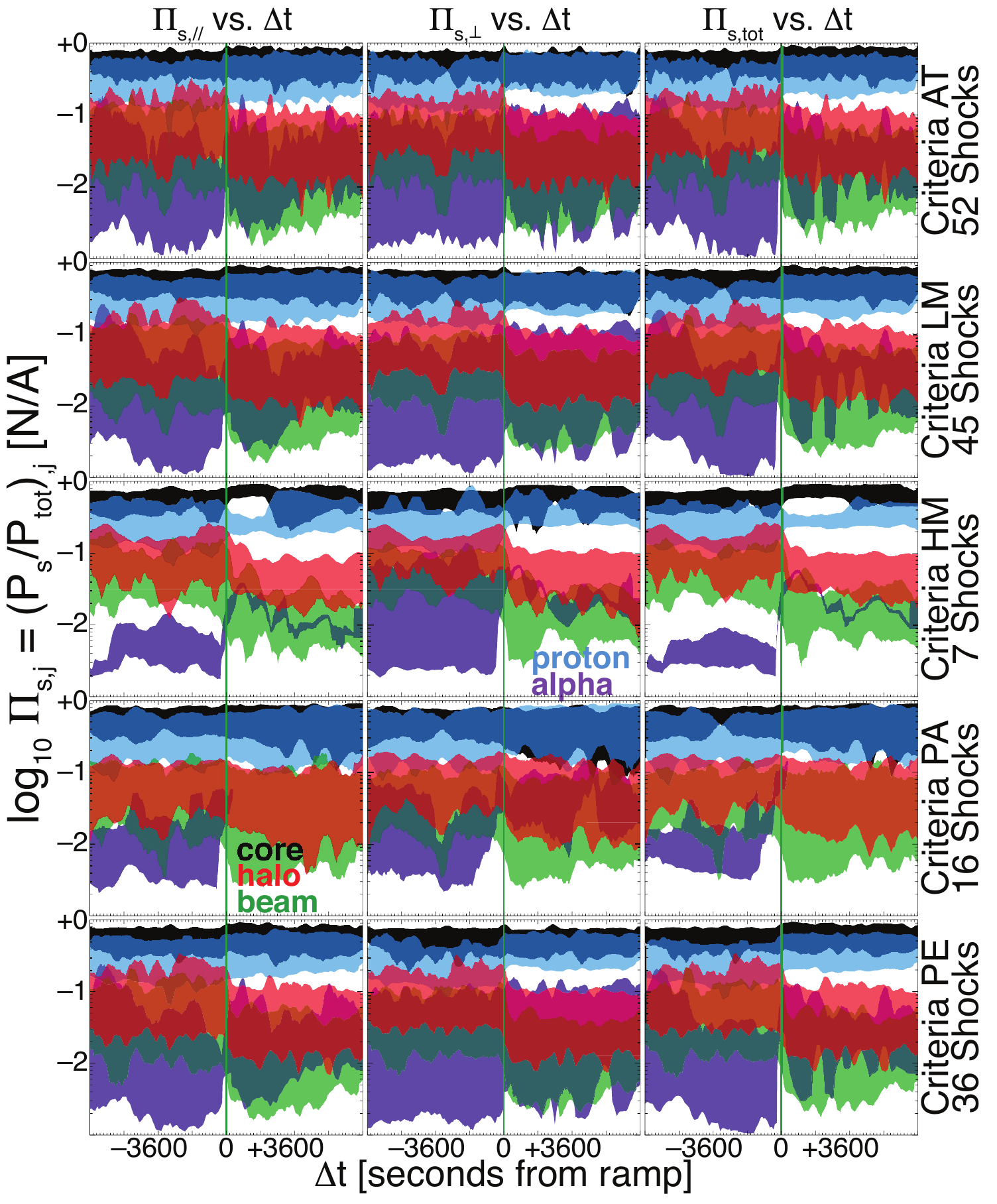}}
    \caption{Shaded regions bounded by $X{\scriptstyle_{5\%}}$ and $X{\scriptstyle_{95\%}}$ (e.g., cyan lines in Figure \ref{fig:Exponents}) from the SEA of the partial thermal pressures, $\Pi{\scriptstyle_{s, j}}$ (see Appendix \ref{app:Definitions} for parameter definitions), for each of the following particle populations:  core electrons (black regions), halo electrons (red regions), beam/strahl electrons (green regions), protons (cyan regions), and alpha-particles (purple/violet regions).  Again, unlike previous SEA plots, these are not normalized to an upstream median value.  The vertical green line indicates the ramp center time.  All panels have uniform horizontal and vertical axis ranges.}
    \label{fig:SEAPercentilesOnly}
\end{figure}

\indent  To illustrate that the energy partitions shown as running medians in Figure \ref{fig:SEAMedianOnly} can be characteristic of most of the data, Figure \ref{fig:SEAPercentilesOnly} shows the same parameters for the same populations but as shaded regions bounded by $X{\scriptstyle_{5\%}}$ and $X{\scriptstyle_{95\%}}$.  What is immediately clear is that again $\Pi{\scriptstyle_{ec, j}}$ and $\Pi{\scriptstyle_{p, j}}$ dominate at all times except for some transient excursions to large values for $\Pi{\scriptstyle_{eh, j}}$.  Both $\Pi{\scriptstyle_{eb, j}}$ and $\Pi{\scriptstyle_{\alpha, j}}$ have relatively large spreads in the range between percentiles, but they still follow the same trends illustrated by the running median.  That is, the suprathermal electron fractional thermal energy density decreases while both ion species increase.  The abrupt end in $\Pi{\scriptstyle_{\alpha, j}}$ data for \textit{Criteria PA} shocks in both Figures \ref{fig:SEAMedianOnly} and \ref{fig:SEAPercentilesOnly} are due to low statistics owing to difficulties in finding high quality fits for the alpha-particle peak \citep[e.g., see SWE fit flag requirements in][]{wilsoniii18b}.

\indent  It is also clear that \textit{Criteria HM} shocks have the largest separation between $\Pi{\scriptstyle_{ec, j}}$ and $\Pi{\scriptstyle_{p, j}}$ for all time periods compared to other selection criteria.  The downstream running median values for \textit{Criteria LM} and \textit{Criteria PE} shocks oscillate out of phase but this is not directly reflected in $X{\scriptstyle_{5\%}}$--$X{\scriptstyle_{95\%}}$.  \textit{Criteria PA} shocks exhibit the most overlap in $X{\scriptstyle_{5\%}}$--$X{\scriptstyle_{95\%}}$ and \textit{Criteria HM} shocks have the least.  In both the running $\tilde{X}$ and $X{\scriptstyle_{5\%}}$--$X{\scriptstyle_{95\%}}$, $\Pi{\scriptstyle_{eh, j}}$ shows a very clear decrease across the shock except for \textit{Criteria PA} shocks where the change in $X{\scriptstyle_{5\%}}$ and $X{\scriptstyle_{95\%}}$ is more difficult to observe.  The abrupt drop in $\Pi{\scriptstyle_{eh, j}}$ and $\Pi{\scriptstyle_{eb, j}}$ across shocks is likely related to the drop in $n{\scriptstyle_{eh}} / n{\scriptstyle_{eff}}$ and $n{\scriptstyle_{eb}} / n{\scriptstyle_{eff}}$ across the shocks because the associated temperatures show slight increases across the shock (see additional SEA plots of $T{\scriptstyle_{s, j}}$ found in \citet[][]{wilsoniii20b}).

\phantomsection   
\section{Discussion}  \label{sec:Discussion}

\indent  Superposed epoch analysis (SEA) of the fit results show that the general trend of the normalized (to the upstream median) exponents for all electron components increases across the shock ramp for all selection criteria\footnote{Except for \textit{Criteria UP} and \textit{Criteria DN}, of course}.  That is, the suprathermal electron tails are steeper and the core electrons at low velocities are flatter, relative to a Maxwellian, and steeper at higher energies.  The normalized density, $n{\scriptstyle_{s}} / n{\scriptstyle_{eff}}$, SEA plots show a general increase (relative to the upstream median) across the shock ramp for the core, but decrease for both halo and beam/strahl, for all selection criteria.  That is, only the fraction of core electrons increases across the shock.  Finally, the ratio of the partial-to-total pressure, $\Pi{\scriptstyle_{s, j}}$, for the protons and alpha-particles increase across the shock, while the ratio for all electron populations decrease.

\begin{figure}
  \centering
    {\includegraphics[trim = 0mm 0mm 0mm 0mm, clip, width=80mm]{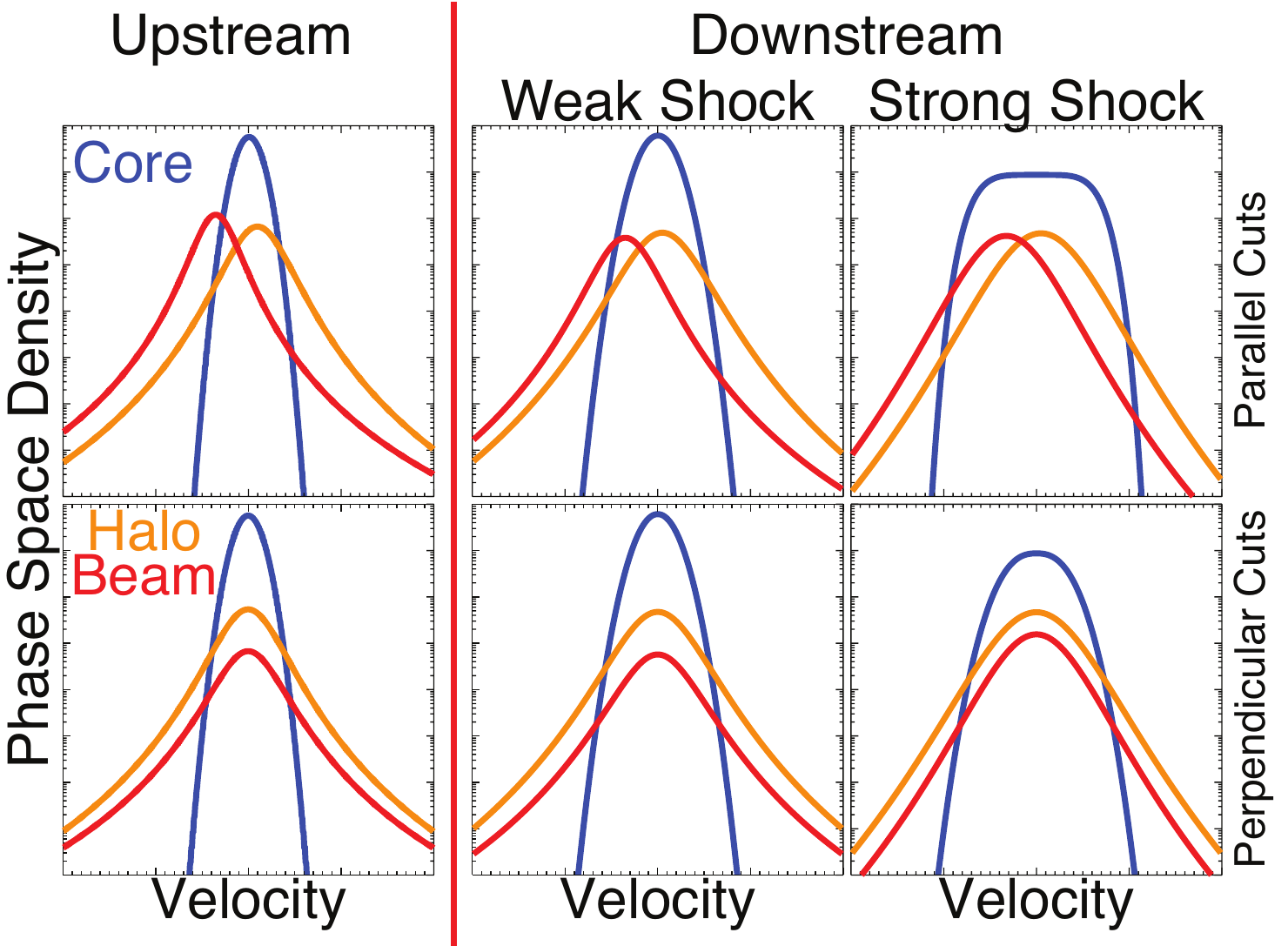}}
    \caption{A cartoon example of the evolution of an electron VDF across the shock ramp showing differences in the downstream one-dimensional cuts between weak and strong shocks.  The color-coded lines indicate the different electron populations of the core (blue), halo (orange), and beam/strahl (red).  The top(bottom) row show cuts parallel(perpendicular) to the local quasi-static magnetic field.}
    \label{fig:CartoonVDF}
\end{figure}

\indent  An illustrative example of the possible electron VDF evolution for the weakest and strongest shocks is qualitatively shown in Figure \ref{fig:CartoonVDF}.  The electron VDF starts as a narrow peaked distribution with hard tails and evolves into a an almost box-like distribution with weaker, soft tails.  Thus, the energy density becomes consolidated into the core population.  It is clear that stronger shocks have a different downstream profile than weaker shocks.  The strong shock example is indicative of the strongest shocks in this study and similar to observations downstream of the terrestrial bow shock.  The larger exponents, in all three electron components, for the strong shock example produces a flatter peak in the core and steeper slopes in the halo and beam/strahl.  Although the number density of the downstream weak shock example is lower than for the strong shock example, the phase space density peak is almost an order of magnitude larger due to the smaller exponents and thermal speeds.  Potential reasons for the change in profile are discussed later.

\indent  The majority of the thermal energy density is held in the core electrons with $\Pi{\scriptstyle_{ec, j}}$ $\sim$ 25--92\% (i.e., fraction of the total pressure) and the protons taking up most of the rest with $\Pi{\scriptstyle_{p, j}}$ $\sim$ 13--72\% for \textit{Criteria AT}.  For $\sim$95\% of the suprathermal electron and alpha-particle velocity moments, none individually satisfy $\Pi{\scriptstyle_{s, j}}$ $>$ 18\%.  That is, the partition of thermal energy density is completely dominated by the core electrons and protons while the suprathermal electrons and alpha-particles serve as mostly minor contributors.  Further, the magnitude of the change in partial pressures, $\widetilde{\Delta \Pi}{\scriptstyle_{s, j}}$, across the 52 IP shocks examined herein (shown as $X{\scriptstyle_{min}}$ $\lesssim$ $X$ $\lesssim$ $X{\scriptstyle_{max}}$) satisfies following:
\begin{itemize}[itemsep=0pt,parsep=0pt,topsep=0pt]
  \item  1.7\% $\lesssim$ $\widetilde{\Delta \Pi}{\scriptstyle_{ec, j}}$ $\lesssim$ 55.8\%;
  \item  0.6\% $\lesssim$ $\widetilde{\Delta \Pi}{\scriptstyle_{eh, j}}$ $\lesssim$ 35.9\%;
  \item  0.3\% $\lesssim$ $\widetilde{\Delta \Pi}{\scriptstyle_{eb, j}}$ $\lesssim$ 16.5\%;
  \item  1.8\% $\lesssim$ $\widetilde{\Delta \Pi}{\scriptstyle_{eff, j}}$ $\lesssim$ 66.7\%;
  \item  1.8\% $\lesssim$ $\widetilde{\Delta \Pi}{\scriptstyle_{p, j}}$ $\lesssim$ 66.7\%; and
  \item  0.07\% $\lesssim$ $\widetilde{\Delta \Pi}{\scriptstyle_{\alpha, j}}$ $\lesssim$ 10.8\%.
\end{itemize}
\noindent  Therefore, the core electrons and protons both carry the largest $\Pi{\scriptstyle_{s, j}}$ and they tend to experience the largest $\widetilde{\Delta \Pi}{\scriptstyle_{s, j}}$.

\indent  Of the three minor partial pressure populations (i.e., halo, beam/strahl, and alpha-particles), the halo electrons consistently dominate in the upstream region, especially $\Pi{\scriptstyle_{eh, \parallel}}$.  In the downstream, $\Pi{\scriptstyle_{eb, \perp}}$ and $\Pi{\scriptstyle_{\alpha, \perp}}$ can be comparable to $\Pi{\scriptstyle_{eh, \perp}}$.  Interestingly, the $\Pi{\scriptstyle_{ec, j}}$ and $\Pi{\scriptstyle_{p, j}}$ often vary out of phase with each other suggesting a partial thermal pressure balance.  This is more weakly reflected in the $\Pi{\scriptstyle_{\alpha, j}}$ variations relative to the $\Pi{\scriptstyle_{eh, j}}$ and $\Pi{\scriptstyle_{eb, j}}$ values.

\indent  Unexpectedly, it was found that the change in pressure of the electrons could be comparable to or larger than the protons.  That is, the ratio of these changes for the 52 IP shocks examined herein (shown as $X{\scriptstyle_{min}}$ $\lesssim$ $X$ $\lesssim$ $X{\scriptstyle_{max}}$, $\tilde{X}$) satisfies following:
\begin{itemize}[itemsep=0pt,parsep=0pt,topsep=0pt]
  \item  16.9\% $\lesssim$ $\widetilde{\Delta P}{\scriptstyle_{ec, j}}$/$\widetilde{\Delta P}{\scriptstyle_{p, j}}$ $\lesssim$ 391\%, $\sim$101\%;
  \item  1.11\% $\lesssim$ $\widetilde{\Delta P}{\scriptstyle_{eh, j}}$/$\widetilde{\Delta P}{\scriptstyle_{p, j}}$ $\lesssim$ 41.3\%, $\sim$5.41\%;
  \item  0.47\% $\lesssim$ $\widetilde{\Delta P}{\scriptstyle_{eb, j}}$/$\widetilde{\Delta P}{\scriptstyle_{p, j}}$ $\lesssim$ 18.3\%, $\sim$2.55\%; and
  \item  19.6\% $\lesssim$ $\widetilde{\Delta P}{\scriptstyle_{eff, j}}$/$\widetilde{\Delta P}{\scriptstyle_{p, j}}$ $\lesssim$ 453\%, $\sim$107\%.
\end{itemize}
\noindent  Therefore, the change in core electron thermal pressure is comparable to or larger than that for the protons.  This is somewhat unexpected because most IP shock observations and theory suggest that the ions should gain more thermal energy than the electrons.  However, previous low Mach number bow shock observations showed that the electrons gained roughly the same fraction of thermal energy as the ions.  The fractional energy density gain by the electrons increases with increasing Mach number in this study but it's unclear if the increase is influenced by difficulty of the \emph{Wind} SWE Faraday cups measuring ion distributiions downstream of the strongest events.

\indent  The nonlinear trend shown in Figure \ref{fig:TempDifferences} between $\widetilde{\Delta T}{\scriptstyle_{s, j}}$ and $\Delta \overline{KE}{\scriptstyle_{shn}}$ is apparent ``by eye'' but as the lower/upper uncertainty bounds on the fit lines illustrate, the trend is moderate at best.  While some previous studies examined a linear relationship between these two parameters, they were primarily focused on observations at the much higher Mach number bow shock.  Further, the fits were performed and shown between these two parameters because they were the best.  Virtually every combination of macroscopic shock parameters was examined against nearly every $\widetilde{\Delta Q}$ but none exhibited a good correlation except between $\Delta \bar{\xi}{\scriptstyle_{shn}}$ and $\widetilde{\Delta \epsilon}{\scriptstyle_{j}}$ (not shown).  However, the good correlation between $\Delta \bar{\xi}{\scriptstyle_{shn}}$ and $\widetilde{\Delta \epsilon}{\scriptstyle_{j}}$ merely shows that as the available free energy increases, the total internal energy increases, which is not surprising.

\indent  The influence of the macroscopic shock parameters is not evident in the current dataset.  In the following, a moderate correlation shows a clear trend but a significant spread in the dependent variable for any given independent variable value (e.g., similar to $\widetilde{\Delta T}{\scriptstyle_{s, j}}$ plots in Figure \ref{fig:TempDifferences}).  There is a moderate, positive correlation between $\Delta \bar{U}{\scriptstyle_{shn}}$ and both $\widetilde{\Delta \Pi}{\scriptstyle_{eb, j}}$ and $\widetilde{\Delta \Pi}{\scriptstyle_{\alpha, j}}$.  There are also moderate, positive correlations between $\widetilde{\Delta \Pi}{\scriptstyle_{\alpha, j}}$ and both $\Delta \overline{KE}{\scriptstyle_{shn}}$ and $\langle M{\scriptstyle_{Te}} \rangle{\scriptstyle_{up}}$.  In summary, none of the energy density ratios showed a clear dependence upon any macroscopic shock parameter.  Nothing was found to show any dependence upon $\theta{\scriptstyle_{Bn}}$.  That is, the change in the fractional thermal energy densities of the five major particle populations appear to be independent of the shock geometry.  Therefore, microscopic instabilities were investigated because they are known to affect each population differently \citep[e.g.,][]{artemyev13j, artemyev14e, artemyev15d, artemyev16b, artemyev17a, artemyev17b, artemyev18a, chang13a, dum78a, dum78b, hughes14a, krall73a, osmane12b, petkaki03a, petkaki06a, sagdeev66}.

\indent  Over 97\% of the \totalshownWHI~VDFs shown in Figure \ref{fig:WHFITests} are unstable to either the whistler heat flux (WHFI) or temperature anisotropy instabilities (WTAI).  Roughly $\sim$1\% and $\sim$14\% of the \totalffpfitsA~VDFs with both ion and electron data are unstable to the firehose and mirror instabilities, respectively, which is over an order of magnitude higher the ambient solar wind estimates \citep[e.g.,][]{chen16b}.  Nearly 30\% of these VDFs satisfy $\tensor*{ \mathcal{T} }{^{e}_{p}}{\scriptstyle_{tot}}$ $\geq$ 3, i.e., the threshold below which ion acoustic waves (IAWs) are predicted to experience heavy Landau damping.  However, only $\sim$5\% of the VDFs satisfied the critical drift velocity (between electrons and ions) threshold necessary to generate current-driven IAWs ignoring temperature gradients which reduce this threshold.  Thus, there is sufficient statistical evidence of unstable VDFs near IP shocks, which is not surprising.

\indent  There are several caveats when interpreting the above instability occurrence rates.  For instance, this work used three electron, one core proton, and one alpha-particle beam populations whereas \citet[][]{chen16b} used one electron, two proton, and one alpha-particle beam populations.  Further, the instability threshold for the WHFI and WTAI corresponds to extremely slow growth rates in excess of 18,000 electron cyclotron periods.  Finally, the IAW thresholds were derived for single, isotropic Maxwellian electron and ion populations.  Thus, these instability occurrence rates are only intended to serve as zeroth order proxies of a fully kinetic treatment that includes non-Maxwellian VDFs and multi-component electron and ion populations.  Despite the caveats and the sometimes unexpectedly high occurrence rates, recent fully kinetic analysis of the ion VDFs in the solar wind finds that most intervals are linearly unstable \citep[e.g.,][]{klein18a, klein19a}.

\indent  The examination of whistler and acoustic instabilities is driven by their difference in resonance energy ranges.  The former tend to scatter with the suprathermal electrons affecting both $\mathcal{A}{\scriptstyle_{s}}$ and $\kappa{\scriptstyle_{s}}$ \citep[e.g.,][]{chang13a, gary11a, hughes14a, saito07}, while the latter strongly scatters the core reducing $\mathcal{A}{\scriptstyle_{ec}}$ and increasing $s{\scriptstyle_{ec}}$ (or $p{\scriptstyle_{ec}}$).  There is some evidence to support the differences in $\mathcal{A}{\scriptstyle_{eh}}$ and $\mathcal{A}{\scriptstyle_{eb}}$ combined with differences in $\kappa{\scriptstyle_{eh}}$ and $\kappa{\scriptstyle_{eb}}$ to suggest the WHFI is present and scattering the suprathermal electrons near these IP shocks.  The core electron exponents ($s{\scriptstyle_{ec}}$, $p{\scriptstyle_{ec}}$, and $q{\scriptstyle_{ec}}$) are consistently larger downstream than upstream and the profile of the downstream core electrons does reach the flattop stage (i.e., $p{\scriptstyle_{ec}}$ $\geq$ 4) in the strongest shocks.  This is consistent with the nonlinear saturation stage of IAWs interacting with electrons \citep[e.g.,][]{dum74a, dum75a, dyrud06, sagdeev66, vedenov63} but it could also be due to quasi-static cross-shock electric fields \citep[e.g.,][]{hull01a, mitchell14a, schwartz88a, schwartz11a, schwartz14a, scudder86c}.  However, the influence of both of these effects should increase with Mach number yet the differences in core electron exponents between low and high Mach number shocks are not statistically significant.  This begs the question of how much energy/momentum goes into increasing the exponent versus increasing the temperature (which does consistently and significantly change across the shocks).

\phantomsection   
\section{Conclusions}  \label{sec:Conclusions}

\indent  The analysis of \totalnfitsall~VDFs observed by the \emph{Wind} spacecraft within $\pm$2 hours of 52 interplanetary (IP) shocks has been presented.  Five primary constituent particle populations were included in the analysis, which are the electron core ($s$ $=$ $ec$), halo ($s$ $=$ $eh$), beam/strahl ($s$ $=$ $eb$), and proton ($s$ $=$ $p$) and alpha-particle ($s$ $=$ $\alpha$) populations.  The analysis revealed that most of the VDFs are at least linearly unstable to one or more instabilities, consistent with a similar conclusion based on different techniques applied to ion data in the ambient solar wind \citep[e.g.,][]{klein18a, klein19a}.  For the weaker shocks, the change in core electron, effective electron, and total ion temperature is positively correlated with the change in kinetic energy across the shock in a way that appears to be nonlinear.

\indent  The evolution of the electron VDF illustrated in Figure \ref{fig:CartoonVDF} shows the qualitative trends observed in this study.  The remaining question is what controls the increase in the core exponent versus temperature and why some strong shocks seem to prefer one or the other while the strongest shocks increase both.  Neither instabilities or quasi-static fields alone appear to be capable of producing the observed changes in the electron VDFs.  Kinetic simulations and theoretical work are required to resolve this discrepancy and are beyond the scope of this study.

\indent  Neither instabilities or quasi-static fields alone are sufficient to explain the evolution of the electron VDFs across the IP shocks.  For instance, some of the stronger shocks showed significant core electron heating but little change in the core exponent (i.e., no additional flattening) while others showed significant increases in the core exponent (i.e., strong flattening) but comparatively weak heating.  Theory and simulations have found that nonlinear stochastic acceleration by ion acoustic waves (i.e., referred to as inelastic collisions herein) can self-consistently generate flattops in the core electron VDFs but it is not clear what fraction of the wave energy goes into increasing the exponent versus increasing the temperature.  Further, previous studies have argued that quasi-static fields are capable of producing flattop VDFs but it's not clear what exponent value should result.  Both of these explanations are plagued by the fact that the downstream core electron exponents are not positively correlated with Mach number or change in kinetic energy, quantities that have been correlated with the amplitude of both the electrostatic waves and quasi-static electric fields in collisionless shocks.  That is, stronger shocks do not consistently generate flatter core electron VDFs.  Addressing such fundamental questions in kinetic theory are planned in future studies.

\indent  In summary, the core electrons and protons carry the most thermal energy density and they also experience the largest changes in thermal energy density across the shocks.  A moderate, positive correlation is found between $\widetilde{\Delta T}{\scriptstyle_{s, j}}$ and $\Delta \overline{KE}{\scriptstyle_{shn}}$, for $s$ $=$ $ec$, $eff$, and $i$.  Weaker correlations are found between some VDF parameters and any other macroscopic shock parameter but nothing is correlated with the shock normal angle.  Surprisingly, the change across the shock in core electron pressure relative to the proton pressure, $\widetilde{\Delta P}{\scriptstyle_{ec, j}}$/$\widetilde{\Delta P}{\scriptstyle_{p, j}}$, was found to be comparable to or larger than unity.  That is, the core electron pressure change was larger than that of the protons in at least 23 of the 52 shocks, for any component.  If only the parallel pressures are examined, the number of shocks increases to 28 of 52.  Future work will examine the details of the kinetic physics involved in collisionless shocks to address these issues.

\acknowledgments
\noindent  The authors thank A.F.- Vi{\~n}as, D.A. Roberts, and R.A. Treumann for useful discussions of basic plasma physics and C. Markwardt for helpful feedback on the usage nuances of his MPFIT software.  The work was supported by the International Space Science Institute's (ISSI) International Teams programme.  L.B.W. was partially supported by \emph{Wind} MO\&DA grants and a Heliophysics Innovation Fund (HIF) grant.  L.-J.C. and S.W. were partially supported by the MMS mission in addition to NASA grants 80NSSC18K1369 and 80NSSC17K0012, NSF grants AGS-1619584 and AGS-1552142, and DOE grant DESC0016278.  D.L.T. was partially supported by NASA grant NNX16AQ50G.  M.L.S. was partially supported by grants NNX14AT26G and NNX13AI75G.  J.C.K. was partially supported by NASA grants NNX14AR78G and 80NSSC18K0986.  D.C. was partially supported by grants NNX17AG30G, GO8-19110A, 80NSSC18K1726, 80NSSC18K1218, and NSF grant 1714658.  S.J.S. was partially supported by the MMS/FIELDS investigation.  C.S.S. was partially supported by NASA grant NNX16AI59G and NSF SHINE grant 1622498.  S.D.B. and C.S.S. were partially supported by NASA grant NNX16AP95G.  S.D.B. acknowledges the support of the Leverhulme Trust Visiting Professor program.  M.P.P. and K.A.G. were supported by Parker Solar Probe instrument funds.

\appendix
\section{Definitions and Notation}  \label{app:Definitions}

\indent  As in Papers I and II, this appendix the symbols and notation used throughout will be defined.  All direction-dependent parameters we use the subscript $j$ to represent the direction where $j$ $=$ $tot$ for the entire distribution, $j$ $=$ $\parallel$ for the the parallel direction, and $j$ $=$ $\perp$ for the perpendicular direction, where parallel/perpendicular is with respect to the quasi-static magnetic field vector, $\mathbf{B}{\scriptstyle_{o}}$ [nT].  The use of the generic subscript $s$ to denote the particle species (e.g., electrons, protons, etc.) or the component of a single particle species (e.g., electron core).  For the electron components, the subscript will be $s$ $=$ $ec$ for the core, $s$ $=$ $eh$ for the halo, $s$ $=$ $eb$ for the beam/strahl, $s$ $=$ $eff$ for the effective, $s$ $=$ $int$ for the integrated (see Paper II for definition), and $s$ $=$ $e$ for the total/entire\footnote{This subscript is used in previous studies and defined here as a reference.  Throughout this manuscript, the use of only $e$ will be exclusively reserved for parameters discussed in previous studies.} population.  Below are the symbol/parameters definitions:
\begin{itemize}[itemsep=0pt,parsep=0pt,topsep=0pt]
  \item[]  \textit{one-variable statistics}
  \begin{itemize}[itemsep=0pt,parsep=0pt,topsep=0pt]
    \item  $X{\scriptstyle_{min}}$ $\equiv$ minimum
    \item  $X{\scriptstyle_{max}}$ $\equiv$ maximum
    \item  $\bar{X}$ $\equiv$ mean
    \item  $\tilde{X}$ $\equiv$ median
    \item  $X{\scriptstyle_{5\%}}$ $\equiv$ 5$^{th}$ percentile
    \item  $X{\scriptstyle_{25\%}}$ $\equiv$ lower quartile
    \item  $X{\scriptstyle_{75\%}}$ $\equiv$ upper quartile
    \item  $X{\scriptstyle_{95\%}}$ $\equiv$ 95$^{th}$ percentile
    \item  $\sigma$ $\equiv$ standard deviation
    \item  $\sigma^{2}$ $\equiv$ variance
  \end{itemize}
  \item[]  \textit{fundamental parameters}
  \begin{itemize}[itemsep=0pt,parsep=0pt,topsep=0pt]
    \item  $\varepsilon{\scriptstyle_{o}}$ $\equiv$ permittivity of free space
    \item  $\mu{\scriptstyle_{o}}$ $\equiv$ permeability of free space
    \item  $c$ $\equiv$ speed of light in vacuum [$km \ s^{-1}$] $=$ $\left( \varepsilon{\scriptstyle_{o}} \ \mu{\scriptstyle_{o}} \right)^{-1/2}$
    \item  $k{\scriptstyle_{B}}$ $\equiv$ the Boltzmann constant [$J \ K^{-1}$]
    \item  $e$ $\equiv$ the fundamental charge [$C$]
  \end{itemize}
  \item[]  \textit{plasma parameters}
  \begin{itemize}[itemsep=0pt,parsep=0pt,topsep=0pt]
    \item  $n{\scriptstyle_{s}}$ $\equiv$ the number density [$cm^{-3}$] of species $s$
    \item  $m{\scriptstyle_{s}}$ $\equiv$ the mass [$kg$] of species $s$
    \item  $Z{\scriptstyle_{s}}$ $\equiv$ the charge state of species $s$
    \item  $q{\scriptstyle_{s}}$ $\equiv$ the charge [$C$] of species $s$ $=$ $Z{\scriptstyle_{s}} \ e$
    \item  $T{\scriptstyle_{s, j}}$ $\equiv$ the scalar temperature [$eV$] of the j$^{th}$ component of species $s$
    \item  $P{\scriptstyle_{s, j}}$ $=$ $n{\scriptstyle_{s}} \ k{\scriptstyle_{B}} \ T{\scriptstyle_{s, j}}$ $\equiv$ the partial thermal pressure [$eV \ cm^{-3}$] of the j$^{th}$ component of species $s$
    \item  $P{\scriptstyle_{t, j}}$ $=$ $\sum_{s} \ P{\scriptstyle_{s, j}}$ $\equiv$ the total pressure [$eV \ cm^{-3}$] of the j$^{th}$ component, summed over all species
    \item  $\tensor*{ \mathcal{T} }{^{s'}_{s}}{\scriptstyle_{j}}$ $=$ $\left(T{\scriptstyle_{s'}}/T{\scriptstyle_{s}}\right){\scriptstyle_{j}}$ $\equiv$ the temperature ratio [N/A] of species $s$ and $s'$ of the j$^{th}$ component
    \item  $\mathcal{A}{\scriptstyle_{s}}$ $=$ $\left(T{\scriptstyle_{\perp}}/T{\scriptstyle_{\parallel}}\right){\scriptstyle_{s}}$ $\equiv$ the temperature anisotropy [N/A] of species $s$
    \item  $V{\scriptstyle_{Ts, j}}$ $\equiv$ the most probable thermal speed [$km \ s^{-1}$] of a one-dimensional velocity distribution (see Equation \ref{eq:params_2})
    \item  $\mathbf{V}{\scriptstyle_{os}}$ $\equiv$ the drift velocity [$km \ s^{-1}$] of species $s$ in the plasma bulk flow rest frame
    \item  $\xi{\scriptstyle_{s, j}}$ $=$ $\tfrac{1}{2} m{\scriptstyle_{s}} \ n{\scriptstyle_{s}} \ V{\scriptstyle_{os, j}}^{2}$ $\equiv$ the ram energy density [$eV \ cm^{-3}$] of the j$^{th}$ component of species $s$ in the plasma bulk flow rest frame
    \item $\epsilon{\scriptstyle_{j}}$ $=$ $\tfrac{ B{\scriptstyle_{o}}^{2} }{2 \ \mu{\scriptstyle_{o}}} \ + \ \sum_{s} \left[ P{\scriptstyle_{s, j}} \ + \xi{\scriptstyle_{s, j}} \right]$ $\equiv$ the total energy density [$eV \ cm^{-3}$] of the j$^{th}$ component of the system in the plasma bulk flow rest frame
    \item $\zeta{\scriptstyle_{s, j}}$ $=$ $\tfrac{ \xi{\scriptstyle_{s, j}} }{ \epsilon{\scriptstyle_{j}} }$ $\equiv$ the ratio of the ram energy density of the j$^{th}$ component of species $s$ to the total energy density [N/A]
    \item $\psi{\scriptstyle_{s, j}}$ $=$ $\tfrac{ P{\scriptstyle_{s, j}} }{ \epsilon{\scriptstyle_{j}} }$ $\equiv$ the ratio of the thermal energy density (partial pressure) of the j$^{th}$ component of species $s$ to the total energy density [N/A]
    \item $\Pi{\scriptstyle_{s, j}}$ $=$ $\tfrac{ P{\scriptstyle_{s, j}} }{ P{\scriptstyle_{t, j}} }$ $\equiv$ the ratio of the partial thermal pressure of the j$^{th}$ component of species $s$ to the total thermal pressure [N/A]
    \item  $C{\scriptstyle_{s}}$ $\equiv$ the sound or ion-acoustic sound speed [$km \ s^{-1}$] \citep[see supplemental PDF file][for definitions]{wilsoniii19k}
    \item  $V{\scriptstyle_{A}}$ $\equiv$ the Alfv\'{e}n speed [$km \ s^{-1}$] \citep[see supplemental PDF file][for definitions]{wilsoniii19k}
    \item  $V{\scriptstyle_{f}}$ $\equiv$ the fast mode speed [$km \ s^{-1}$] \citep[see supplemental PDF file][for definitions]{wilsoniii19k}
    \item  $\Omega{\scriptstyle_{cs}}$ $\equiv$ the angular cyclotron frequency [$rad \ s^{-1}$] (see Equation \ref{eq:params_3})
    \item  $\omega{\scriptstyle_{ps}}$ $\equiv$ the angular plasma frequency [$rad \ s^{-1}$] (see Equation \ref{eq:params_4})
    \item  $\lambda{\scriptstyle_{De}}$ $\equiv$ the electron Debye length [$m$] (see Equation \ref{eq:params_5})
    \item  $\rho{\scriptstyle_{cs}}$ $\equiv$ the thermal gyroradius [$km$] (see Equation \ref{eq:params_6})
    \item  $\lambda{\scriptstyle_{s}}$ $\equiv$ the inertial length [$km$] (see Equation \ref{eq:params_7})
    \item  $\beta{\scriptstyle_{s, j}}$ $\equiv$ the plasma beta [N/A] of the j$^{th}$ component of species $s$ (see Equations \ref{eq:params_8} and \ref{eq:params_9})
    \item  $\kappa{\scriptstyle_{s}}$ $\equiv$ the kappa exponent of species $s$ \citep[e.g., see][for definition in model fit equation]{wilsoniii19a}
    \item  $s{\scriptstyle_{s}}$ $\equiv$ the symmetric self-similar exponent of species $s$ \citep[e.g., see][for definition in model fit equation]{wilsoniii19a}
    \item  $p{\scriptstyle_{s}}$($q{\scriptstyle_{s}}$) $\equiv$ the parallel(perpendicular) asymmetric self-similar exponent of species $s$ \citep[e.g., see][for definition in model fit equation]{wilsoniii19a}
    \item  $\phi{\scriptstyle_{sc}}$ $\equiv$ the scalar, quasi-static spacecraft potential [eV] \citep[e.g.,][]{pulupa14a, scime94a} (see Appendices of Paper I for more details)
    \item  $E{\scriptstyle_{min}}$ $\equiv$ the minimum energy bin midpoint value [eV] of an electrostatic analyzer \citep[e.g., see Appendices in][]{wilsoniii17c, wilsoniii18b}
    \item  $q{\scriptstyle_{e, \parallel}}$ $=$ $\tfrac{m{\scriptstyle_{e}}}{2} \ \int \ d^{3}v \ f{\scriptstyle_{e}}^{\left( mod \right)} v{\scriptstyle_{\parallel}} \ v^{2}$ $\equiv$ the parallel electron heat flux [$\mu W \ m^{-2}$] of the entire electron VDF model, $f{\scriptstyle_{e}}^{\left( mod \right)}$ $=$ $f^{\left( core \right)}$ $+$ $f^{\left( halo \right)}$ $+$ $f^{\left( beam \right)}$
    \item  $q{\scriptstyle_{e o}}$ $=$ $\tfrac{3}{2} \ m{\scriptstyle_{e}} \ n{\scriptstyle_{e}} \ V{\scriptstyle_{Tec, \parallel}}^{3}$ $\equiv$ the free-streaming limit electron heat flux [$\mu W \ m^{-2}$] \citep[e.g.,][]{gary99a}
  \end{itemize}
\end{itemize}

\noindent  Similar to Paper I, the variables that rely upon multiple parameters are given in the following equations:

\begin{subequations}
  \begin{align}
    T{\scriptstyle_{eff, j}} & = \frac{ \sum_{s} n{\scriptstyle_{s}} \ T{\scriptstyle_{s, j}} }{ \sum_{s} n{\scriptstyle_{s}} } \label{eq:params_0} \\
    T{\scriptstyle_{s, tot}} & = \frac{1}{3} \left( T{\scriptstyle_{s, \parallel}} + 2 \ T{\scriptstyle_{s, \perp}} \right) \label{eq:params_1} \\
    V{\scriptstyle_{Ts, j}} & = \sqrt{ \frac{ 2 \ k{\scriptstyle_{B}} \ T{\scriptstyle_{s, j}} }{ m{\scriptstyle_{s}} } } \label{eq:params_2} \\
    \Omega{\scriptstyle_{cs}} & = \frac{ q{\scriptstyle_{s}} \ B{\scriptstyle_{o}} }{ m{\scriptstyle_{s}} } \label{eq:params_3} \\
    \omega{\scriptstyle_{ps}} & = \sqrt{ \frac{ n{\scriptstyle_{s}} \ q{\scriptstyle_{s}}^{2} }{ \varepsilon{\scriptstyle_{o}} \ m{\scriptstyle_{s}} } } \label{eq:params_4} \\
    \lambda{\scriptstyle_{De}} & = \frac{ V{\scriptstyle_{Te, tot}} }{ \sqrt{ 2 } \ \omega{\scriptstyle_{pe}} } = \sqrt{ \frac{ \varepsilon{\scriptstyle_{o}} \ k{\scriptstyle_{B}} \ T{\scriptstyle_{e, tot}} }{ n{\scriptstyle_{e}} \ e^{2} } } \label{eq:params_5} \\
    \rho{\scriptstyle_{cs}} & = \frac{ V{\scriptstyle_{Ts, tot}} }{ \Omega{\scriptstyle_{cs}} } \label{eq:params_6} \\
    \lambda{\scriptstyle_{s}} & = \frac{ c }{ \omega{\scriptstyle_{ps}} } \label{eq:params_7} \\
    \beta{\scriptstyle_{s, j}} & = \frac{ 2 \mu{\scriptstyle_{o}} n{\scriptstyle_{s}} k{\scriptstyle_{B}} T{\scriptstyle_{s, j}} }{ \lvert \mathbf{B}{\scriptstyle_{o}} \rvert^{2} } \label{eq:params_8} \\
    \beta{\scriptstyle_{eff, j}} & = \frac{ 2 \mu{\scriptstyle_{o}} n{\scriptstyle_{eff}} k{\scriptstyle_{B}} T{\scriptstyle_{eff, j}} }{ \lvert \mathbf{B}{\scriptstyle_{o}} \rvert^{2} } \label{eq:params_9} \\
    \intertext{where $n{\scriptstyle_{eff}}$ is defined as:}
    n{\scriptstyle_{eff}} & = \sum_{s} \ n{\scriptstyle_{es}} \label{eq:params_10}
  \end{align}
\end{subequations}

\indent  For the macroscopic shock parameters, the values are averaged over asymptotic regions away from the shock transition region.

\begin{itemize}[itemsep=0pt,parsep=0pt,topsep=0pt]
  \item[]  \textit{shock parameters}
  \begin{itemize}[itemsep=0pt,parsep=0pt,topsep=0pt]
    \item  subscripts $up$ and $dn$ $\equiv$ denote the upstream (i.e., before the shock arrives time-wise at the spacecraft for a forward shock) and downstream (i.e., the shocked region)
    \item  $\langle Q \rangle{\scriptstyle_{j}}$ $\equiv$ the average of parameter $Q$ over the $j^{th}$ shock region, where $j$ $=$ $up$ or $dn$
    \item  $\Delta \bar{Q}$ $=$ $\langle Q \rangle{\scriptstyle_{dn}}$ - $\langle Q \rangle{\scriptstyle_{up}}$ $\equiv$ the change in the asymptotic average of parameter $Q$ over the $j^{th}$ shock region
    \item  $\bar{R}{\scriptstyle_{ns}}$ $=$ $\langle n{\scriptstyle_{s}} \rangle{\scriptstyle_{dn}}$/$\langle n{\scriptstyle_{s}} \rangle{\scriptstyle_{up}}$ $\equiv$ the average shock compression ratio of species $s$
    \item  $\bar{R}{\scriptstyle_{Qs,j}}$ $=$ $\langle Q{\scriptstyle_{s,j}} \rangle{\scriptstyle_{dn}}$/$\langle Q{\scriptstyle_{s,j}} \rangle{\scriptstyle_{up}}$ $\equiv$ the downstream-to-upstream $j^{th}$ component ratio of the asymptotic average of parameter $Q$ of species $s$
    \item  $\Delta \mathcal{Q}{\scriptstyle_{\alpha, \beta}}$ $=$ $Q{\scriptstyle_{\alpha}}$ - $Q{\scriptstyle_{\beta}}$ $\equiv$ the set of all permutations of the difference between all the downstream ($\alpha$) and all the upstream ($\beta$) values
    \item  $\widetilde{\Delta Q}$ $\equiv$ the median of $\Delta \mathcal{Q}{\scriptstyle_{\alpha, \beta}}$, i.e., the median value of all permutations of all differences across the shock of parameter $Q$
    \item  $\mathbf{n}{\scriptstyle_{sh}}$ $\equiv$ the shock normal unit vector [N/A]
    \item  $\theta{\scriptstyle_{Bn}}$ $\equiv$ the shock normal angle\footnote{The acute reference angle between $\langle \mathbf{B}{\scriptstyle_{o}} \rangle{\scriptstyle_{up}}$ and $\mathbf{n}{\scriptstyle_{sh}}$.} [deg]
    \item  $\langle \lvert V{\scriptstyle_{shn}} \rvert \rangle{\scriptstyle_{j}}$ $\equiv$ the $j^{th}$ region average shock normal speed [$km \ s^{-1}$] in the spacecraft frame
    \item  $\langle \lvert U{\scriptstyle_{shn}} \rvert \rangle{\scriptstyle_{j}}$ $\equiv$ the $j^{th}$ region average shock normal speed [$km \ s^{-1}$] in the shock rest frame (i.e., the speed of the flow relative to the shock)
    \item  $\langle \lvert KE{\scriptstyle_{shn}} \rvert \rangle{\scriptstyle_{j}}$ $=$ $\tfrac{1}{2} \ m{\scriptstyle_{p}} \ \langle \lvert U{\scriptstyle_{shn}} \rvert \rangle{\scriptstyle_{j}}^{2}$ $\equiv$ the $j^{th}$ region average shock normal kinetic energy [$eV$] in the shock rest frame
    \item  $\langle \lvert \xi{\scriptstyle_{shn}} \rvert \rangle{\scriptstyle_{j}}$ $=$ $\tfrac{1}{2} \ m{\scriptstyle_{p}} \ \langle n{\scriptstyle_{p}} \rangle{\scriptstyle_{j}} \  \langle \lvert U{\scriptstyle_{shn}} \rvert \rangle{\scriptstyle_{j}}^{2}$ $\equiv$ the $j^{th}$ region average shock normal kinetic energy density [$eV \ cm^{-3}$] in the shock rest frame
    \item  $\langle M{\scriptstyle_{A}} \rangle{\scriptstyle_{j}}$ $=$ $\langle \lvert U{\scriptstyle_{shn}} \rvert \rangle{\scriptstyle_{j}} / \langle V{\scriptstyle_{A}} \rangle{\scriptstyle_{j}}$ $\equiv$ the $j^{th}$ region average Alfv\'{e}nic Mach number [N/A]
    \item  $\langle M{\scriptstyle_{f}} \rangle{\scriptstyle_{j}}$ $=$ $\langle \lvert U{\scriptstyle_{shn}} \rvert \rangle{\scriptstyle_{j}} / \langle V{\scriptstyle_{f}} \rangle{\scriptstyle_{j}}$ $\equiv$ the $j^{th}$ region average fast mode Mach number [N/A]
    \item  $\langle M{\scriptstyle_{Te}} \rangle{\scriptstyle_{j}}$ $=$ $\langle \lvert U{\scriptstyle_{shn}} \rvert \rangle{\scriptstyle_{j}} / \langle V{\scriptstyle_{Teff, tot}} \rangle{\scriptstyle_{j}}$ $\equiv$ the $j^{th}$ region average electron thermal Mach number [N/A]
    \item  $M{\scriptstyle_{cr}}$ $\equiv$ the first critical Mach number [N/A]
  \end{itemize}
\end{itemize}

\indent  For brevity the percent difference between one-variable statistics values for two different selection criteria are defined here.  The percent difference between parameters satisfying \textit{Criteria DN} and \textit{Criteria UP} is defined as $\Delta Q{\scriptstyle_{d2u}}$ $=$ $\left( Q{\scriptstyle_{dn}} - Q{\scriptstyle_{up}} \right)/Q{\scriptstyle_{up}} \times 100\%$, where $Q$ is any one-variable statistic value.  Similarly, the percent difference between \textit{Criteria HM} and \textit{Criteria LM} is defined as $\Delta Q{\scriptstyle_{h2l}}$ $=$ $\left( Q{\scriptstyle_{HM}} - Q{\scriptstyle_{LM}} \right)/Q{\scriptstyle_{LM}} \times 100\%$ and that between \textit{Criteria PE} and \textit{Criteria PA} is $\Delta Q{\scriptstyle_{\perp 2 \parallel}}$ $=$ $\left( Q{\scriptstyle_{PE}} - Q{\scriptstyle_{PA}} \right)/Q{\scriptstyle_{PA}} \times 100\%$.

\indent  As in Paper II, integrated velocity moments refer to the velocity moments calculated by integrating over the entire model function, $f{\scriptstyle_{e}}^{\left( mod \right)}$ $=$ $f^{\left( core \right)}$ $+$ $f^{\left( halo \right)}$ $+$ $f^{\left( beam \right)}$, rather than the fit values from the components.  The integrated moments are only calculated for VDFs with stable solutions for all three components using the Simpson's $\tfrac{1}{3}$ Rule algorithm.  The integrals are calculated in the core electron rest frame, thus the only relevant heat flux component is the parallel, $q{\scriptstyle_{e, \parallel}}$, because the suprathermal electrons have no finite perpendicular drift velocities (e.g., see Paper I).  For further details on the integrated velocity moments, see Paper II.

\noindent  These definitions are used throughout.

\clearpage
\phantomsection   
\section{Extra Statistics}  \label{app:ExtraStatistics}

\indent  In this section some extra statistics are presented in tabular form in Tables \ref{tab:CEnergyDensRatios} and \ref{tab:PressureRatios}.

\startlongtable  
\begin{deluxetable*}{| l | c | c | c | c | c | c | c | c |}
  \tabletypesize{\footnotesize}    
  \tablecaption{Component Thermal Pressure-to-Total Energy Density Ratios as Percentages \label{tab:CEnergyDensRatios}}
  \tablehead{\colhead{Ratios} & \colhead{$X{\scriptstyle_{min}}$}\tablenotemark{a} & \colhead{$X{\scriptstyle_{5\%}}$}\tablenotemark{b} & \colhead{$X{\scriptstyle_{25\%}}$}\tablenotemark{c} & \colhead{$\bar{X}$}\tablenotemark{d} & \colhead{$\tilde{X}$}\tablenotemark{e} & \colhead{$X{\scriptstyle_{75\%}}$}\tablenotemark{f} & \colhead{$X{\scriptstyle_{95\%}}$}\tablenotemark{g} & \colhead{$X{\scriptstyle_{max}}$}\tablenotemark{h}}
  \startdata
  \multicolumn{9}{ |c| }{\textit{Criteria AT: \totalnfitsall~VDFs}} \\
  \hline
  $\psi{\scriptstyle_{ec, \parallel}}$       & 4.54  & 15.6 & 25.3 & 35.6 & 32.2 & 46.9 & 61.1 & 94.9  \\
  $\psi{\scriptstyle_{ec, \perp}}$           & 4.81  & 11.5 & 23.3 & 35.1 & 32.5 & 48.3 & 61.9 & 95.1  \\
  $\psi{\scriptstyle_{ec, tot}}$             & 6.02  & 14.3 & 23.8 & 34.0 & 31.4 & 45.3 & 57.5 & 95.1  \\
  \hline
  $\psi{\scriptstyle_{eh, \parallel}}$       & 0.007 & 0.72 & 1.70 & 4.10 & 3.27 & 5.32 & 11.0 & 71.1  \\
  $\psi{\scriptstyle_{eh, \perp}}$           & 0.02  & 0.78 & 1.81 & 4.40 & 3.52 & 5.67 & 11.6 & 72.4  \\
  $\psi{\scriptstyle_{eh, tot}}$             & 0.02  & 0.77 & 1.74 & 4.12 & 3.34 & 5.39 & 10.5 & 72.0  \\
  \hline
  $\psi{\scriptstyle_{eb, \parallel}}$       & 0.002 & 0.18 & 0.69 & 2.11 & 1.55 & 2.92 & 5.88 & 17.3  \\
  $\psi{\scriptstyle_{eb, \perp}}$           & 0.001 & 0.23 & 0.69 & 1.97 & 1.50 & 2.70 & 5.40 & 14.3  \\
  $\psi{\scriptstyle_{eb, tot}}$             & 0.001 & 0.21 & 0.70 & 1.93 & 1.47 & 2.68 & 5.11 & 13.7  \\
  \hline
  $\psi{\scriptstyle_{eff, \parallel}}$      & 0.09  & 18.0 & 29.4 & 40.9 & 37.5 & 53.8 & 70.5 & 99.4  \\
  $\psi{\scriptstyle_{eff, \perp}}$          & 0.10  & 13.2 & 27.2 & 40.5 & 37.8 & 54.4 & 72.0 & 99.4  \\
  $\psi{\scriptstyle_{eff, tot}}$            & 0.06  & 16.5 & 27.8 & 39.2 & 36.2 & 50.8 & 66.7 & 99.4  \\
  \hline
  $\psi{\scriptstyle_{p, \parallel}}$        & 0.45  & 5.96 & 12.4 & 23.9 & 21.2 & 32.1 & 51.4 & 92.1  \\
  $\psi{\scriptstyle_{p, \perp}}$            & 0.61  & 6.92 & 14.6 & 23.6 & 20.9 & 30.6 & 49.0 & 89.4  \\
  $\psi{\scriptstyle_{p, tot}}$              & 0.61  & 6.11 & 13.4 & 21.9 & 19.6 & 27.8 & 45.8 & 81.7  \\
  \hline
  $\psi{\scriptstyle_{\alpha, \parallel}}$   & 0.05  & 0.15 & 0.28 & 1.28 & 0.58 & 1.22 & 5.23 & 20.3  \\
  $\psi{\scriptstyle_{\alpha, \perp}}$       & 0.01  & 0.12 & 0.27 & 1.74 & 0.81 & 2.47 & 6.07 & 25.9  \\
  $\psi{\scriptstyle_{\alpha, tot}}$         & 0.03  & 0.14 & 0.26 & 1.37 & 0.59 & 1.92 & 4.91 & 13.6  \\
  \hline
  \multicolumn{9}{ |c| }{\textit{Criteria UP: \totalnfitsups~VDFs}} \\
  \hline
  $\psi{\scriptstyle_{ec, \parallel}}$       & 4.54  & 14.9 & 26.9 & 37.4 & 33.8 & 49.5 & 62.9 & 86.0  \\
  $\psi{\scriptstyle_{ec, \perp}}$           & 3.88  & 11.9 & 26.7 & 38.7 & 37.0 & 51.8 & 64.6 & 85.6  \\
  $\psi{\scriptstyle_{ec, tot}}$             & 3.34  & 11.6 & 25.9 & 35.9 & 33.6 & 47.4 & 58.9 & 85.7  \\
  \hline
  $\psi{\scriptstyle_{eh, \parallel}}$       & 0.02  & 0.98 & 2.96 & 5.54 & 4.62 & 6.92 & 13.3 & 71.1  \\
  $\psi{\scriptstyle_{eh, \perp}}$           & 0.10  & 1.18 & 3.39 & 6.14 & 5.04 & 7.62 & 15.3 & 72.4  \\
  $\psi{\scriptstyle_{eh, tot}}$             & 0.09  & 1.06 & 3.09 & 5.56 & 4.63 & 6.98 & 13.4 & 72.0  \\
  \hline
  $\psi{\scriptstyle_{eb, \parallel}}$       & 0.002 & 0.47 & 1.54 & 3.08 & 2.62 & 4.15 & 7.02 & 17.3  \\
  $\psi{\scriptstyle_{eb, \perp}}$           & 0.04  & 0.53 & 1.52 & 2.89 & 2.55 & 3.75 & 6.47 & 14.3  \\
  $\psi{\scriptstyle_{eb, tot}}$             & 0.02  & 0.50 & 1.48 & 2.78 & 2.49 & 3.65 & 6.03 & 13.5  \\
  \hline
  $\psi{\scriptstyle_{eff, \parallel}}$      & 0.12  & 19.8 & 33.0 & 44.7 & 41.5 & 55.6 & 74.2 & 96.8  \\
  $\psi{\scriptstyle_{eff, \perp}}$          & 0.20  & 17.5 & 33.9 & 46.4 & 45.2 & 58.9 & 76.4 & 96.8  \\
  $\psi{\scriptstyle_{eff, tot}}$            & 0.06  & 18.1 & 32.1 & 43.1 & 41.7 & 52.9 & 68.4 & 96.8  \\
  \hline
  $\psi{\scriptstyle_{p, \parallel}}$        & 0.45  & 5.11 & 11.2 & 21.3 & 18.6 & 29.5 & 45.9 & 81.4  \\
  $\psi{\scriptstyle_{p, \perp}}$            & 0.92  & 5.21 & 11.9 & 19.1 & 17.0 & 23.9 & 42.3 & 82.2  \\
  $\psi{\scriptstyle_{p, tot}}$              & 0.80  & 4.96 & 11.2 & 18.4 & 16.7 & 24.2 & 38.6 & 73.2  \\
  \hline
  $\psi{\scriptstyle_{\alpha, \parallel}}$   & 0.05  & 0.11 & 0.24 & 0.97 & 0.36 & 0.82 & 4.33 & 13.8  \\
  $\psi{\scriptstyle_{\alpha, \perp}}$       & 0.05  & 0.10 & 0.21 & 1.13 & 0.43 & 1.28 & 4.96 & 16.8  \\
  $\psi{\scriptstyle_{\alpha, tot}}$         & 0.06  & 0.11 & 0.22 & 0.81 & 0.35 & 0.80 & 3.04 & 10.4  \\
  \hline
  \multicolumn{9}{ |c| }{\textit{Criteria DN: \totalnfitsdns~VDFs}} \\
  \hline
  $\psi{\scriptstyle_{ec, \parallel}}$       & 4.91  & 16.0 & 24.3 & 34.3 & 31.0 & 43.7 & 58.7 & 94.9  \\
  $\psi{\scriptstyle_{ec, \perp}}$           & 0.48  & 11.4 & 21.3 & 32.5 & 29.7 & 41.8 & 59.8 & 95.1  \\
  $\psi{\scriptstyle_{ec, tot}}$             & 0.60  & 14.7 & 22.6 & 32.6 & 30.1 & 42.5 & 56.2 & 95.1  \\
  \hline
  $\psi{\scriptstyle_{eh, \parallel}}$       & 0.007 & 0.63 & 1.39 & 3.06 & 2.41 & 4.06 & 7.45 & 25.0  \\
  $\psi{\scriptstyle_{eh, \perp}}$           & 0.02  & 0.67 & 1.46 & 3.15 & 2.57 & 4.18 & 7.67 & 22.6  \\
  $\psi{\scriptstyle_{eh, tot}}$             & 0.02  & 0.68 & 1.43 & 3.08 & 2.50 & 4.11 & 7.47 & 23.0  \\
  \hline
  $\psi{\scriptstyle_{eb, \parallel}}$       & 0.002 & 0.14 & 0.49 & 1.38 & 1.00 & 1.84 & 3.88 & 16.0  \\
  $\psi{\scriptstyle_{eb, \perp}}$           & 0.001 & 0.16 & 0.50 & 1.27 & 0.93 & 1.72 & 3.39 & 14.1  \\
  $\psi{\scriptstyle_{eb, tot}}$             & 0.001 & 0.16 & 0.51 & 1.29 & 0.94 & 1.73 & 3.34 & 13.7  \\
  \hline
  $\psi{\scriptstyle_{eff, \parallel}}$      & 0.09  & 17.7 & 26.8 & 38.0 & 34.6 & 49.7 & 64.7 & 99.4  \\
  $\psi{\scriptstyle_{eff, \perp}}$          & 0.10  & 12.3 & 23.9 & 36.2 & 33.5 & 46.2 & 65.1 & 99.4  \\
  $\psi{\scriptstyle_{eff, tot}}$            & 0.09  & 16.2 & 25.2 & 36.3 & 33.8 & 47.6 & 62.2 & 99.4  \\
  \hline
  $\psi{\scriptstyle_{p, \parallel}}$        & 1.05  & 7.04 & 14.0 & 26.1 & 23.5 & 35.6 & 54.4 & 92.1  \\
  $\psi{\scriptstyle_{p, \perp}}$            & 0.61  & 9.93 & 17.5 & 27.1 & 23.6 & 34.3 & 52.5 & 89.4  \\
  $\psi{\scriptstyle_{p, tot}}$              & 0.61  & 10.1 & 16.1 & 25.0 & 21.8 & 32.0 & 48.7 & 81.7  \\
  \hline
  $\psi{\scriptstyle_{\alpha, \parallel}}$   & 0.10  & 0.27 & 0.57 & 1.90 & 0.90 & 2.56 & 6.31 & 20.3  \\
  $\psi{\scriptstyle_{\alpha, \perp}}$       & 0.01  & 0.19 & 0.69 & 2.62 & 1.96 & 3.84 & 6.85 & 25.9  \\
  $\psi{\scriptstyle_{\alpha, tot}}$         & 0.03  & 0.30 & 0.73 & 2.46 & 2.05 & 3.51 & 5.80 & 13.6  \\
  \hline
  \enddata
  \tablenotetext{a}{minimum}
  \tablenotetext{b}{5$^{th}$ percentile}
  \tablenotetext{c}{25$^{th}$ percentile}
  \tablenotetext{d}{mean}
  \tablenotetext{e}{median}
  \tablenotetext{f}{75$^{th}$ percentile}
  \tablenotetext{g}{95$^{th}$ percentile}
  \tablenotetext{h}{maximum}
  \tablecomments{All values represented as percentages.  For symbol definitions, see Appendix \ref{app:Definitions}.}
\vspace{-20pt}
\end{deluxetable*}

\startlongtable  
\begin{deluxetable*}{| l | c | c | c | c | c | c | c | c |}
  \tabletypesize{\footnotesize}    
  \tablecaption{Component-to-Total Thermal Pressure Ratios as Percentages \label{tab:PressureRatios}}
  \tablehead{\colhead{Ratios} & \colhead{$X{\scriptstyle_{min}}$} & \colhead{$X{\scriptstyle_{5\%}}$} & \colhead{$X{\scriptstyle_{25\%}}$} & \colhead{$\bar{X}$} & \colhead{$\tilde{X}$} & \colhead{$X{\scriptstyle_{75\%}}$} & \colhead{$X{\scriptstyle_{95\%}}$} & \colhead{$X{\scriptstyle_{max}}$}}
  \startdata
  \multicolumn{9}{ |c| }{\textit{Criteria AT: \totalnfitsall~VDFs}} \\
  \hline
  $\Pi{\scriptstyle_{ec, \parallel}}$       & 7.00  & 30.3 & 45.3 & 60.4 & 61.8 & 73.8 & 91.6 & 99.9  \\
  $\Pi{\scriptstyle_{ec, \perp}}$           & 4.36  & 25.9 & 43.7 & 56.6 & 57.9 & 69.0 & 87.2 & 99.9  \\
  $\Pi{\scriptstyle_{ec, tot}}$             & 6.67  & 31.9 & 47.6 & 60.9 & 61.2 & 73.7 & 91.1 & 99.9  \\
  \hline
  $\Pi{\scriptstyle_{eh, \parallel}}$       & 0.02  & 1.36 & 3.12 & 7.01 & 5.39 & 9.12 & 16.3 & 99.4  \\
  $\Pi{\scriptstyle_{eh, \perp}}$           & 0.05  & 1.39 & 3.26 & 7.27 & 5.71 & 9.45 & 17.0 & 96.8  \\
  $\Pi{\scriptstyle_{eh, tot}}$             & 0.14  & 1.50 & 3.43 & 7.55 & 5.91 & 9.85 & 17.5 & 97.8  \\
  \hline
  $\Pi{\scriptstyle_{eb, \parallel}}$       & 0.003 & 0.30 & 1.16 & 3.84 & 2.59 & 5.21 & 11.0 & 80.6  \\
  $\Pi{\scriptstyle_{eb, \perp}}$           & 0.003 & 0.36 & 1.15 & 3.39 & 2.45 & 4.62 & 8.99 & 73.2  \\
  $\Pi{\scriptstyle_{eb, tot}}$             & 0.004 & 0.39 & 1.28 & 3.69 & 2.66 & 5.03 & 9.44 & 75.8  \\
  \hline
  $\Pi{\scriptstyle_{eff, \parallel}}$      & 0.15  & 33.8 & 49.6 & 63.4 & 65.9 & 78.1 & 86.6 & 98.9  \\
  $\Pi{\scriptstyle_{eff, \perp}}$          & 0.23  & 29.4 & 48.4 & 61.4 & 65.0 & 75.0 & 85.4 & 97.2  \\
  $\Pi{\scriptstyle_{eff, tot}}$            & 0.19  & 35.3 & 52.6 & 63.7 & 66.0 & 76.2 & 85.4 & 97.8  \\
  \hline
  $\Pi{\scriptstyle_{p, \parallel}}$        & 1.09  & 13.3 & 21.6 & 36.4 & 34.0 & 50.3 & 66.3 & 99.9  \\
  $\Pi{\scriptstyle_{p, \perp}}$            & 2.84  & 14.3 & 24.1 & 37.8 & 33.9 & 50.4 & 71.9 & 99.8  \\
  $\Pi{\scriptstyle_{p, tot}}$              & 2.17  & 14.4 & 23.3 & 36.0 & 33.2 & 47.3 & 64.6 & 99.8  \\
  \hline
  $\Pi{\scriptstyle_{\alpha, \parallel}}$   & 0.07  & 0.19 & 0.35 & 1.92 & 0.82 & 2.07 & 7.17 & 28.8  \\
  $\Pi{\scriptstyle_{\alpha, \perp}}$       & 0.07  & 0.17 & 0.36 & 2.98 & 1.45 & 4.03 & 10.7 & 43.4  \\
  $\Pi{\scriptstyle_{\alpha, tot}}$         & 0.08  & 0.20 & 0.37 & 2.26 & 0.93 & 3.16 & 8.60 & 25.4  \\
  \hline
  \multicolumn{9}{ |c| }{\textit{Criteria UP: \totalnfitsups~VDFs}} \\
  \hline
  $\Pi{\scriptstyle_{ec, \parallel}}$       & 11.2  & 30.9 & 47.0 & 59.3 & 62.1 & 71.7 & 84.1 & 98.4  \\
  $\Pi{\scriptstyle_{ec, \perp}}$           & 5.98  & 27.1 & 50.9 & 60.0 & 63.3 & 70.8 & 82.3 & 98.0  \\
  $\Pi{\scriptstyle_{ec, tot}}$             & 13.4  & 31.4 & 50.8 & 60.5 & 63.3 & 71.4 & 83.4 & 97.8  \\
  \hline
  $\Pi{\scriptstyle_{eh, \parallel}}$       & 0.04  & 1.74 & 4.58 & 9.15 & 7.47 & 11.2 & 21.8 & 99.4  \\
  $\Pi{\scriptstyle_{eh, \perp}}$           & 0.22  & 1.98 & 5.43 & 9.87 & 8.32 & 11.9 & 21.9 & 96.8  \\
  $\Pi{\scriptstyle_{eh, tot}}$             & 0.20  & 1.95 & 5.18 & 9.86 & 8.19 & 11.9 & 22.4 & 97.8  \\
  \hline
  $\Pi{\scriptstyle_{eb, \parallel}}$       & 0.003 & 0.76 & 2.16 & 5.46 & 4.18 & 7.10 & 14.6 & 80.6  \\
  $\Pi{\scriptstyle_{eb, \perp}}$           & 0.04  & 0.85 & 2.27 & 4.88 & 4.09 & 6.25 & 11.4 & 73.2  \\
  $\Pi{\scriptstyle_{eb, tot}}$             & 0.03  & 0.85 & 2.30 & 5.19 & 4.23 & 6.70 & 13.0 & 75.8  \\
  \hline
  $\Pi{\scriptstyle_{eff, \parallel}}$      & 0.84  & 39.9 & 56.3 & 68.2 & 73.0 & 81.3 & 87.9 & 98.9  \\
  $\Pi{\scriptstyle_{eff, \perp}}$          & 1.49  & 36.6 & 62.9 & 70.0 & 73.3 & 82.0 & 88.0 & 97.2  \\
  $\Pi{\scriptstyle_{eff, tot}}$            & 1.97  & 39.3 & 62.0 & 69.9 & 73.2 & 81.8 & 87.9 & 97.8  \\
  \hline
  $\Pi{\scriptstyle_{p, \parallel}}$        & 1.09  & 11.9 & 18.3 & 32.0 & 27.0 & 43.6 & 61.9 & 99.2  \\
  $\Pi{\scriptstyle_{p, \perp}}$            & 2.84  & 11.6 & 17.9 & 29.9 & 26.0 & 35.6 & 67.2 & 99.1  \\
  $\Pi{\scriptstyle_{p, tot}}$              & 2.17  & 11.9 & 18.1 & 30.1 & 26.7 & 37.8 & 62.1 & 98.8  \\
  \hline
  $\Pi{\scriptstyle_{\alpha, \parallel}}$   & 0.07  & 0.15 & 0.28 & 1.50 & 0.51 & 1.44 & 5.81 & 28.8  \\
  $\Pi{\scriptstyle_{\alpha, \perp}}$       & 0.08  & 0.15 & 0.26 & 1.99 & 0.62 & 2.26 & 8.53 & 36.6  \\
  $\Pi{\scriptstyle_{\alpha, tot}}$         & 0.08  & 0.16 & 0.28 & 1.47 & 0.54 & 1.54 & 5.56 & 25.4  \\
  \hline
  \multicolumn{9}{ |c| }{\textit{Criteria DN: \totalnfitsdns~VDFs}} \\
  \hline
  $\Pi{\scriptstyle_{ec, \parallel}}$       & 7.00  & 30.0 & 44.2 & 61.3 & 61.4 & 77.6 & 94.3 & 99.9  \\
  $\Pi{\scriptstyle_{ec, \perp}}$           & 4.36  & 24.9 & 40.5 & 54.1 & 52.1 & 65.1 & 89.3 & 99.9  \\
  $\Pi{\scriptstyle_{ec, tot}}$             & 6.67  & 32.2 & 45.4 & 61.2 & 59.4 & 77.2 & 93.5 & 99.9  \\
  \hline
  $\Pi{\scriptstyle_{eh, \parallel}}$       & 0.02  & 1.19 & 2.61 & 5.45 & 4.20 & 7.08 & 13.1 & 90.7  \\
  $\Pi{\scriptstyle_{eh, \perp}}$           & 0.05  & 1.13 & 2.64 & 5.40 & 4.29 & 6.78 & 13.4 & 94.4  \\
  $\Pi{\scriptstyle_{eh, tot}}$             & 0.01  & 1.27 & 2.88 & 5.88 & 4.61 & 7.57 & 14.1 & 93.1  \\
  \hline
  $\Pi{\scriptstyle_{eb, \parallel}}$       & 0.005 & 0.22 & 0.81 & 2.62 & 1.72 & 3.56 & 8.01 & 64.2  \\
  $\Pi{\scriptstyle_{eb, \perp}}$           & 0.003 & 0.27 & 0.81 & 2.27 & 1.58 & 3.09 & 6.27 & 21.8  \\
  $\Pi{\scriptstyle_{eb, tot}}$             & 0.004 & 0.28 & 0.92 & 2.57 & 1.81 & 3.50 & 7.07 & 51.1  \\
  \hline
  $\Pi{\scriptstyle_{eff, \parallel}}$      & 0.15  & 31.2 & 45.8 & 59.5 & 60.7 & 74.0 & 84.7 & 95.4  \\
  $\Pi{\scriptstyle_{eff, \perp}}$          & 0.23  & 26.9 & 43.6 & 54.8 & 54.7 & 68.5 & 79.3 & 93.7  \\
  $\Pi{\scriptstyle_{eff, tot}}$            & 0.19  & 33.2 & 46.8 & 58.4 & 59.6 & 71.0 & 80.7 & 94.0  \\
  \hline
  $\Pi{\scriptstyle_{p, \parallel}}$        & 4.55  & 15.2 & 25.6 & 40.2 & 38.7 & 54.0 & 68.6 & 99.9  \\
  $\Pi{\scriptstyle_{p, \perp}}$            & 6.01  & 20.2 & 30.8 & 44.0 & 42.9 & 55.4 & 73.5 & 99.8  \\
  $\Pi{\scriptstyle_{p, tot}}$              & 5.99  & 19.3 & 28.6 & 41.2 & 39.9 & 53.0 & 66.5 & 99.8  \\
  \hline
  $\Pi{\scriptstyle_{\alpha, \parallel}}$   & 0.13  & 0.36 & 0.77 & 2.74 & 1.27 & 3.91 & 8.58 & 28.6  \\
  $\Pi{\scriptstyle_{\alpha, \perp}}$       & 0.07  & 0.26 & 1.38 & 4.40 & 3.40 & 5.92 & 12.2 & 43.4  \\
  $\Pi{\scriptstyle_{\alpha, tot}}$         & 0.23  & 0.45 & 0.99 & 3.81 & 3.21 & 5.21 & 10.1 & 23.2  \\
  \hline
  \enddata
  \tablecomments{All values represented as percentages.  The header symbols are the same as in Table \ref{tab:CEnergyDensRatios}.  For symbol definitions, see Appendix \ref{app:Definitions}.}
\vspace{-20pt}
\end{deluxetable*}


\end{document}